%% file: zmain.tex
\documentclass[twocolumn,trackchanges]{aastex62}
\usepackage[section]{placeins}
\usepackage{longtable}
\usepackage{hyperref}
\usepackage{url}
\usepackage{amsmath}

\newcommand{\ptde}{$P(\textrm{TDE})$}

\usepackage{enumitem}
\usepackage{fontawesome}

\usepackage[hang,flushmargin]{footmisc} 
\usepackage{url}

\input{affil}
\shorttitle{FLEET-TDE}
\shortauthors{Gomez et al.}

\begin{document}

\title{Identifying Tidal Disruption Events with an Expansion of the FLEET Machine Learning Algorithm}

\correspondingauthor{Sebastian Gomez}
\email{sgomez@stsci.edu}

\author[0000-0001-6395-6702]{Sebastian Gomez}
\STScI

\author[0000-0002-5814-4061]{V. Ashley Villar}
\PSUa\PSUb\PSUc

\author[0000-0002-9392-9681]{Edo Berger}
\CfA\IAIFI

\author[0000-0003-3703-5154]{Suvi Gezari}
\STScI

\author[0000-0002-3859-8074]{Sjoert van Velzen}
\Leiden

\author[0000-0002-2555-3192]{Matt Nicholl}
\Birmingham

\author[0000-0003-0526-2248]{Peter K. Blanchard}
\CIERA

\author[0000-0002-8297-2473]{Kate. D. Alexander}
\Einstein\CIERA

\begin{abstract}

We present an expansion of FLEET, a machine learning algorithm originally designed to find superluminous supernovae in time-domain surveys, now optimized to select for transients that are most likely to be tidal disruption events (TDEs). FLEET is based on a random forest algorithm trained on both the light curves and host galaxy information of 4,779 spectroscopically classified transients, including 45 TDEs. We find that for transients with a probability of being a TDE, \ptde$>0.5$, we can successfully recover TDEs with a $\approx 40$\% completeness and a $\approx 30$\% purity when using the first 20 days of light curve photometry, or a similar completeness and $\approx 50$\% purity when including 40 days of photometry, an improvement of almost two orders of magnitude compared to random selection. Alternatively, we can recover TDEs with a maximum purity of $\approx 80$\% when considering only transients with \ptde$>0.8$. We find that the most relevant features for differentiating TDEs from other transients are the normalized host separation, and the light curve $(g-r)$ color during peak. Additionally, we use FLEET to produce a list of the 39 most likely TDE candidates discovered by the Zwicky Transient Facility that remain currently unclassified. We explore the use of FLEET for future time-domain surveys such as the Legacy Survey of Space and Time on the Vera C.~Rubin Observatory (\textit{Rubin}) and the \textit{Nancy Grace Roman Space Telescope} (\textit{Roman}) High Latitude Time Domain Survey. We simulate the \textit{Rubin} and \textit{Roman} survey strategies and estimate that $\sim 10^4$ well observed TDEs could be discovered every year by \textit{Rubin}, and $\sim 200$ TDEs per year by \textit{Roman}. Finally, we run FLEET on the TDEs in our \textit{Rubin} survey simulation and find that we can recover $\sim 30$\% of those at a redshift $z < 0.5$ with \ptde$> 0.5$. This translates to $\sim 3,000$ TDEs per year that FLEET could uncover from the \textit{Rubin} stream. Demonstrating that we can begin running FLEET on \textit{Rubin} photometry and discovering interesting transients as soon as the survey begins. FLEET is provided as an open source package on GitHub \href{https://github.com/gmzsebastian/FLEET}{\faGithub}.

\end{abstract}

\keywords{black hole physics -- supernovae: general -- methods: statistical -- surveys}

\section{Introduction}\label{sec:intro}

Tidal disruption events (TDEs) occur when a star gets too close to a supermassive black hole and is torn apart by the tidal forces of the black hole \citep{hills75,Rees88}. Following this disruption, about half the stellar debris is expected to return towards the black hole and circularize into an accretion disk, beginning a phase of accretion when a bright optical transient can be observed \citep{Gezari09, Guillochon09}. To date, about 70 TDEs have been discovered across the electromagnetic spectrum, with a wide variation of observational features \citep{Auchettl17, Mockler19, Velzen20, Velzen20_TDEs, Gezari21, Hammerstein22, Nicholl22}. Some optically discovered TDEs exhibit hydrogen and helium emission, while others only helium \citep{Gezari12, Arcavi14}, and some TDEs had nitrogen and oxygen lines detected \citep{Blagorodnova19, Leloudas19}. \cite{Velzen20} defined three classes of TDEs: TDE-H (hydrogen only), TDE-He (helium only) and TDE-H+He (Bowen lines in combination with H and/or He), and at least one TDE has evolved from TDE-H+He to TDE-He \citep{Nicholl19_17eqx}. Some TDEs show X-ray emission in excess of their optical luminosity, others are X-ray dim \citep{Holoien16,Auchettl17}, while others alternate between these states \citep{Gezari17}. Additionally, radio observations suggest that some TDEs produce fast outflows, including relativistic jets, while others do not \citep{Zauderer11, Bower13, Velzen13, Alexander16, Velzen16, Alexander20, Cendes22}.

Finding more TDEs rapidly and efficiently will aid towards a better understanding of their nature, origin, and evolution. The challenge in finding more TDEs lies in the fact that these are rare, representing only $\sim 0.5$\% of all spectroscopically classified transients from magnitude limited surveys. Nevertheless, the field of time domain astronomy is growing at an increasing rate, with surveys like Zwicky Transient Facility (ZTF) discovering thousands of transients a month \citep{Bellm19}. With existing resources, it is impossible to spectroscopically classify more than $\sim 10$\% of all optical transients discovered. This will become even more challenging when the Legacy Survey of Space and Time on the Vera C.~Rubin Observatory \textit{Rubin} commences in 2024, increasing the transient discovery rate by $\sim 2$ orders of magnitude \citep{Ivezic19}.

Machine learning (ML) algorithms can be used to select promising TDE candidates for more efficient spectroscopic follow-up. Some general purpose ML classifiers that attempt to predict the classes of optical transients already exist, but none are trained to classify TDEs using real observational data. For example, {\tt SuperRAENN} \citep{Villar20} and {\tt Superphot} \citep{Villar19,Hosseinzadeh20} have been trained on real data from the Pan-STARRS1 Medium Deep Survey (PS1 MDS) and use a recurrent autoencoder neural network and a random forest algorithm, respectively, to predict the classes of five types of supernovae (SNe Ia, Ibc, II, IIn, and SLSN-I). However, TDEs are not one of the supported classes in these classifiers, since the PS1 MDS training set only included 2 spectroscopically classified TDEs. The {\tt RAPID} \citep{Muthukrishna19} and {\tt Avocado} \citep{Boone19} classifiers on the other hand are trained to distinguish among at least 12 different transient classes, including SNe, TDEs, and stellar flares, but were trained on simulated data from the Photometric LSST Astronomical Time-series Classification project (PLAsTiCC; \citealt{Kessler19}) based on observations of 11 TDEs, and their effectiveness on real data is yet to be verified. Lastly, the Automatic Learning for the Rapid Classification of Events (ALeRCE) broker \citep{Sanchez21} uses a two-stage random forest algorithm that first classifies the general nature of the transient as either an AGN, SN, Variable Star, Asteroid, or Bogus, then proceeds to refine the class into 15 different types, but TDEs are not one of these classes.

Given the absence of robust TDEs photometric classifiers, here we present a new version of FLEET (\textit{Finding Luminous and Exotic Extragalactic Transients}), a ML algorithm originally developed to target follow-up of Type-I superluminous supernovae (SLSNe) based on their photometric classification \citep{Gomez20}. Over the past two years of operations, we have proven the efficacy of FLEET in finding SLSNe and surpassed our predictions of performance, where we managed to achieve a peak purity of $\approx 80$\% and discovered 21 of the 50 SLSNe found worldwide since we deployed FLEET in November 2019 (Gomez et al. 2022, jointly submitted). Here, we use a similar approach of using a transient's light curve and host galaxy information to target TDEs, without focusing on the classification of other transients, with the aim of obtaining the purest possible sample of TDEs. Unlike existing algorithms, FLEET is trained on real data, and makes use of both light curve information and contextual host galaxy parameters to predict the likelihood of a transient being a TDE, without the need to know its redshift. FLEET is designed to be fast and capable of classifying thousands of transients in a matter of a few hours on a personal computer.

ML algorithms such as FLEET will prove not only useful but necessary as more transient surveys come online. We estimate the expected number of TDEs that could be detected by both \textit{Rubin} and the Nancy Grace Roman Space Telescope (\textit{Roman}; \citealt{Spergel15}), and explore the possibility of using FLEET to target transients from these surveys and maximize our efficiency in recovering them.

The structure of the paper is as follows. In \S\ref{sec:data} we outline the sources of data used for training FLEET, in \S\ref{sec:fleet} we describe the underlying algorithm, in \S\ref{sec:selection} we explore possible selection effects in FLEET, in \S\ref{sec:surveys} we describe the use of FLEET in the ZTF, \textit{Rubin}, and \textit{Roman} time-domain surveys, and finally conclude in \S\ref{sec:conclusions}. FLEET is provided as a Python package on Github\footnote{\url{https://github.com/gmzsebastian/FLEET}} and Zenodo \citep{Gomez20_FLEET}, as well as included in the Python Package Index under the name {\tt fleet-pipe}.

\section{Data}\label{sec:data}

\begin{figure}
    \begin{center}
    \centering
    {\includegraphics[width=\columnwidth]{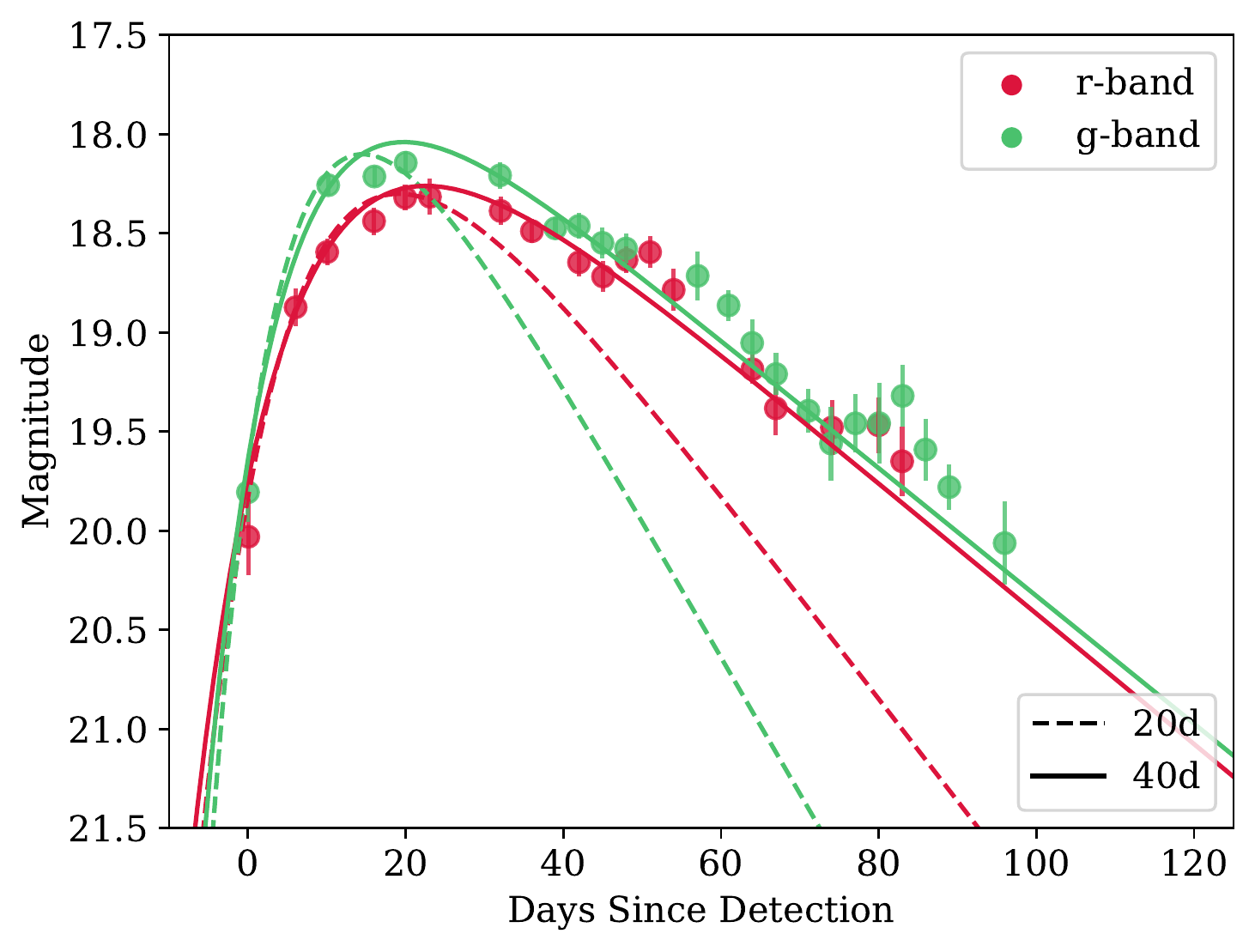}}
    \caption{Light curves of the TDE AT\,2019ehz fit with the model described in Equation~\ref{eq:lc}. The dashed lines show the fit using only data up to 20 days after detection (with a fixed value of $A=0.6$), while the solid lines are the result of fitting the data up to 40 days after detection (with $A$ as a free parameter). The former is part of our rapid classifier, while the latter is part of our late-time classifier.
    \label{fig:2019ehz}}
    \end{center}
\end{figure}

To build a training sample for FLEET, we gathered a list of all spectroscopically classified transients from the Transient Name Server (TNS)\footnote{\label{ref:tns}\url{https://www.wis-tns.org/}}. We then restrict the sample to only include transients with at least 2 $g$-band and 2 $r$-band photometric detections, to have enough data points to fit a model to their light curves. We further restrict the sample to only include and transients that lie within the footprint of the Pan-STARRS1 $3\pi$ (PS1/$3\pi$) survey \citep{Chambers18}, for the purpose of identifying their host galaxies. The resulting sample is composed of 4779 transients, with the following distinct labels from the TNS: 2983 SN\,Ia, 749 SN\,II, 187 SLSN-I, 157 SN\,IIn, 143 CV, 105 SN\,Ic, 89 SN\,IIP, 80 SN\,Ib, 68 SN\,IIb, 52 SLSN-II, 45 TDE, 35 SN\,Ic-BL, 26 SN\,Ibc, 23 AGN, 19 SN\,Ibn, 18 variable star. We provide a list the 45 TDEs used for this training sample in Table~A.\ref{tab:TDEs}.

The light curves of all the transients used for training are from the Zwicky Transient Facility (ZTF; \citealt{Bellm19}), and were obtained from the Automatic Learning for the Rapid Classification of Events (ALeRCE) broker \citep{Forster20}. While we only require 4 detections for a transient to be included in the training set, 90\% of them have at least 8 detections, and 50\% have at least 22 detections in either $g$- or $r$-band. We extract host galaxy information from the PS1/$3\pi$ \citep{Chambers18} and Sloan Digital Sky Survey (SDSS) catalogs \citep{Alam15,Ahumada19}. Finally, we correct all photometry for Galactic extinction using the \cite{Schlafly11} estimates of $E(B-V)$ and the \cite{Barbary16} implementation of the \cite{Cardelli89} extinction law.

\begin{figure}
    \begin{center}
    \centering
    {\includegraphics[width=\columnwidth]{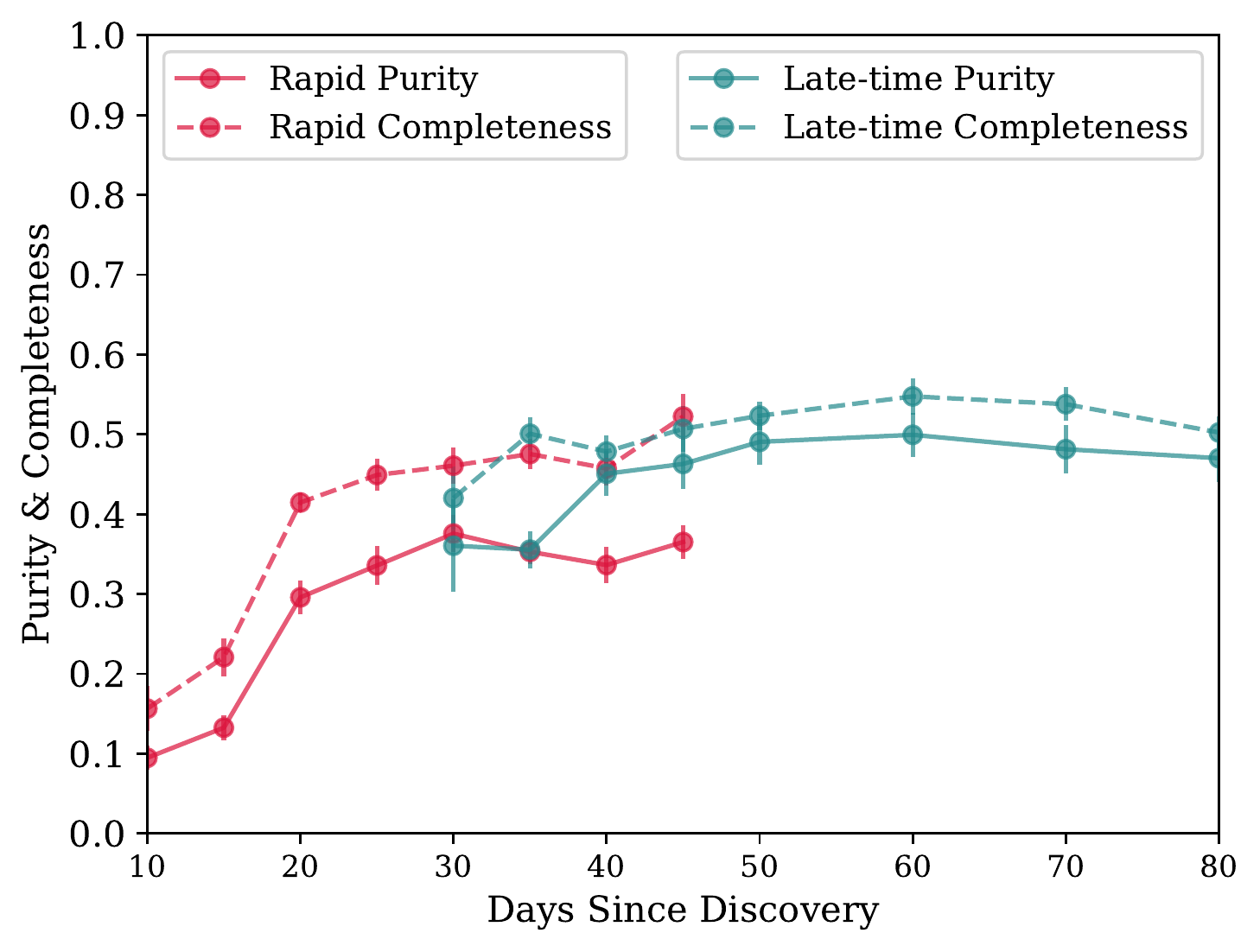}}
    \caption{Purity and completeness as a function of days of photometry included in the models. The dashed lines represent the completeness, while the solid lines represent purity. The rapid model is shown in red, while the late-time model is shown in teal. \label{fig:days_tde}}
    \end{center}
\end{figure}

\section{FLEET Algorithm}\label{sec:fleet}

\begin{figure}
\begin{center}
\centering
{\includegraphics[width=\columnwidth]{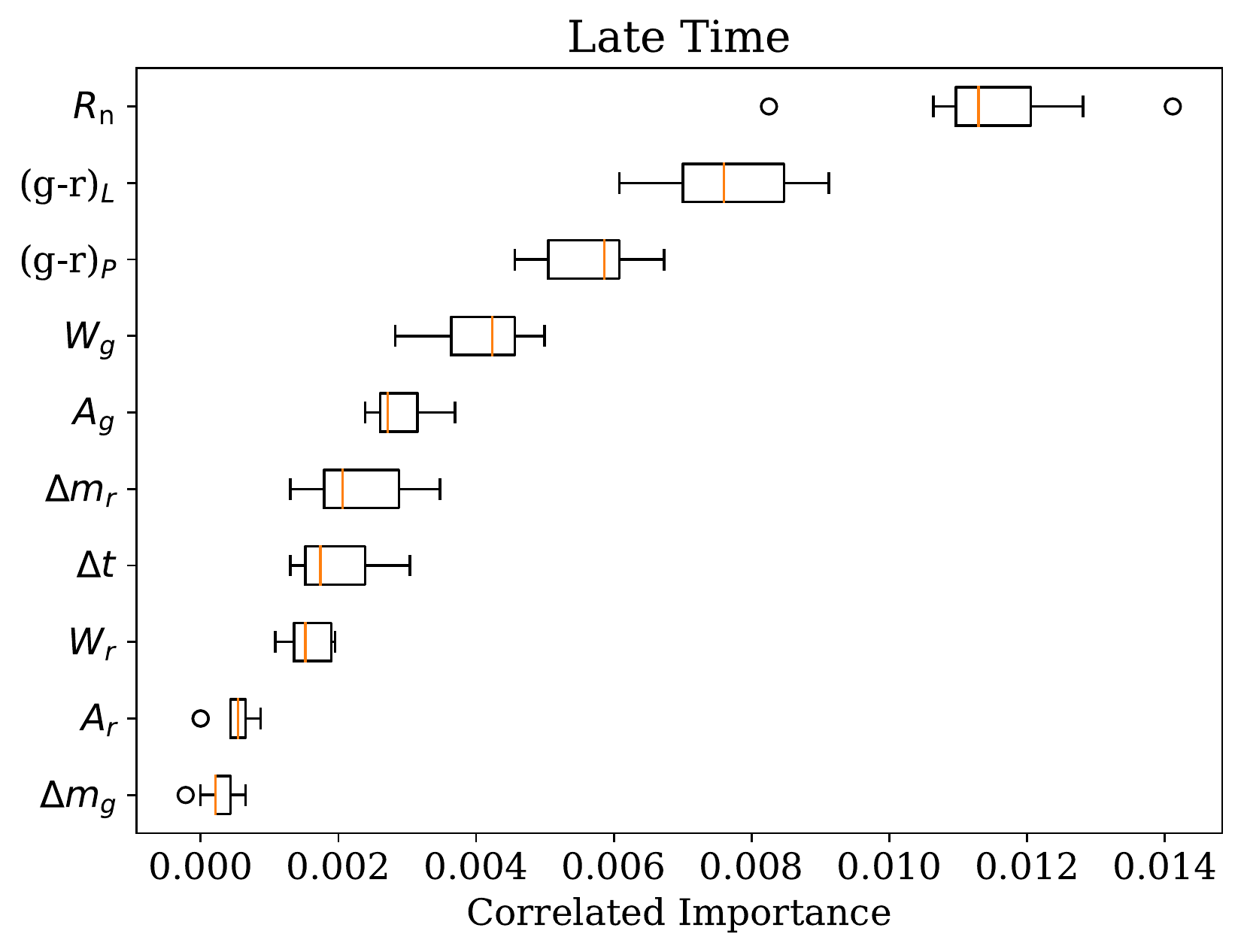}
\includegraphics[width=\columnwidth]{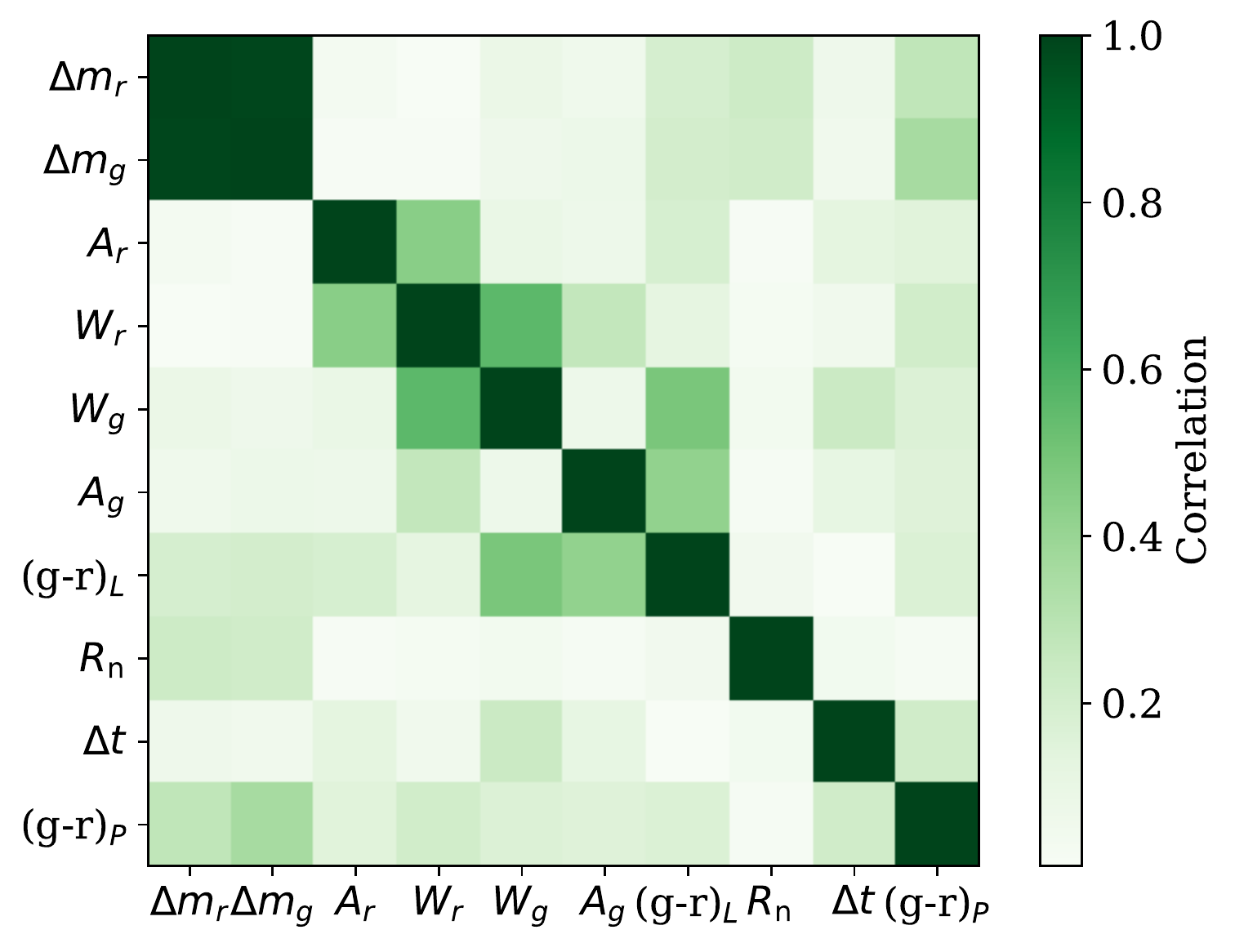}}
\caption{\textit{Top}: Correlated importance for the features used in the late-time version of the classifier, we find that the host separation $R_n$, and transient colors $(g-r)_P$ and $(g-r)_L$ appear to be the most important features for predictive accuracy. \textit{Bottom} : Correlation matrix for the same features. Except for $W_g$ and $W_r$, most features show a low correlation. \label{fig:correlation}}
\end{center}
\end{figure}

FLEET was originally designed to target SLSNe from transient alert streams. Here, we expand the capabilities of FLEET and train it to find TDEs. For a full description of the algorithm see \cite{Gomez20}, a short description follows.

FLEET uses the random forest (RF) implementation from the {\tt scikit-learn} Python package \citep{Pedregosa12} to predict the probability of a transient to be a TDE, \ptde. The training set of all classified transients described in \S\ref{sec:data} is unbalanced, meaning not every class of transient has the same number of events. To prevent the classifier from being biased towards predicting the more common classes and to improve predictive accuracy, we oversample all transients to have a population size equal to that of the largest transient class (i.e., 2983 SNe Ia) using the Synthetic Minority Over-sampling Technique (SMOTE; \citealt{Chawla02}). In \cite{Gomez20} we determined the optimal grouping of transients classes to be: Nuclear, SLSN-I, SLSN-II, SN II, SN IIb, SN IIn, SN Ia, SN Ibc, and Star. The only difference in class labels between the TDE classifier and the original one, is that the former ``Nuclear" class has been split into distinct ``AGN" and ``TDE" classes. Allowing the classifier to group transients into these distinct classes improves its overall accuracy. Nevertheless, we are not concerned with the individual classifications of other transients since they are eventually compressed into a binary ``TDE" versus ``not-TDE" classification. The uncertainties for all predictions from the RF algorithm presented here represent the 1$\sigma$ scatter of 25 different realizations of each model, generated using a different initial random seed.

We query a $1'$ region of the PS1 and SDSS catalogs around each transient to identify their most likely host galaxy. First, we assign a probability of being a galaxy to every object in the field to rule out stars. We estimate this probability using a custom k-nearest-neighbors algorithm trained on data from the Canada-France-Hawaii Telescope Legacy Survey (CFHTLS; \citealt{Hudelot12}), which uses the {\tt CLASS\_STAR} classifier flag in {\tt SExtractor} to separate stars from galaxies, relying on a multi-layer feed-forward neural network \citep{Bertin96}. Then, we calculate the probability of chance coincidence $P_{\rm cc}$ of every galaxy using the method of \cite{Bloom02} described in \cite{Berger10} and select the galaxy with the lowest $P_{\rm cc}$ as the host galaxy of the transient. This method is computationally fast, and accurate for $\sim 95$\% of the associated transients, determined from manual vetting of the training set. This is slightly lower than more complex algorithms, for example the success rate of $\sim 99$\% reported by the DELIGHT algorithm \citep{Foster22}, and $\sim 97$\% from the GHOST classifier \citep{Gagliano21}. The issue of host galaxy association is less ambiguous for TDEs, which thanks to their nuclear nature have a mean $P_{\rm cc} \sim 10^{-3}$, and their host association is therefore effectively always successful. This is compared to the transients in our overall training set, which have a mean $P_{\rm cc} \sim 2.5\times10^{-3}$, which is less confident but still implies a very high likelihood of association.

Since we are motivated by rapid classification, we fit the light curves of transients with a simple, computationally efficient model:
\begin{equation}
\label{eq:lc}
m = e^{W (t - \phi)} - A \times W (t - \phi) + m_0,
\end{equation}
where $W$ is the effective width of the light curve, $A$ modifies the decline time relative to the rise time, $m_0$ is the peak magnitude, and $\phi$ is a phase offset relative to the time of the first observation. We provide two versions of the model. One is aimed at finding TDEs at early times, using only the first 20 days of data after discovery, and has a fixed value of $A = 0.6$ (the mean $A$ for all transients with full light curves). The second model uses the first 40 days of data and includes $A$ as a free parameter. This model is able to more confidently predict a transient class but at the expense of triggering follow-up at a later phase. In Figure~\ref{fig:2019ehz} we show an example of both models fit to a TDE light curve, with independent fits to the $g$- and $r$-band light curves.

\begin{figure}
\begin{center}
\centering
{\includegraphics[width=\columnwidth]{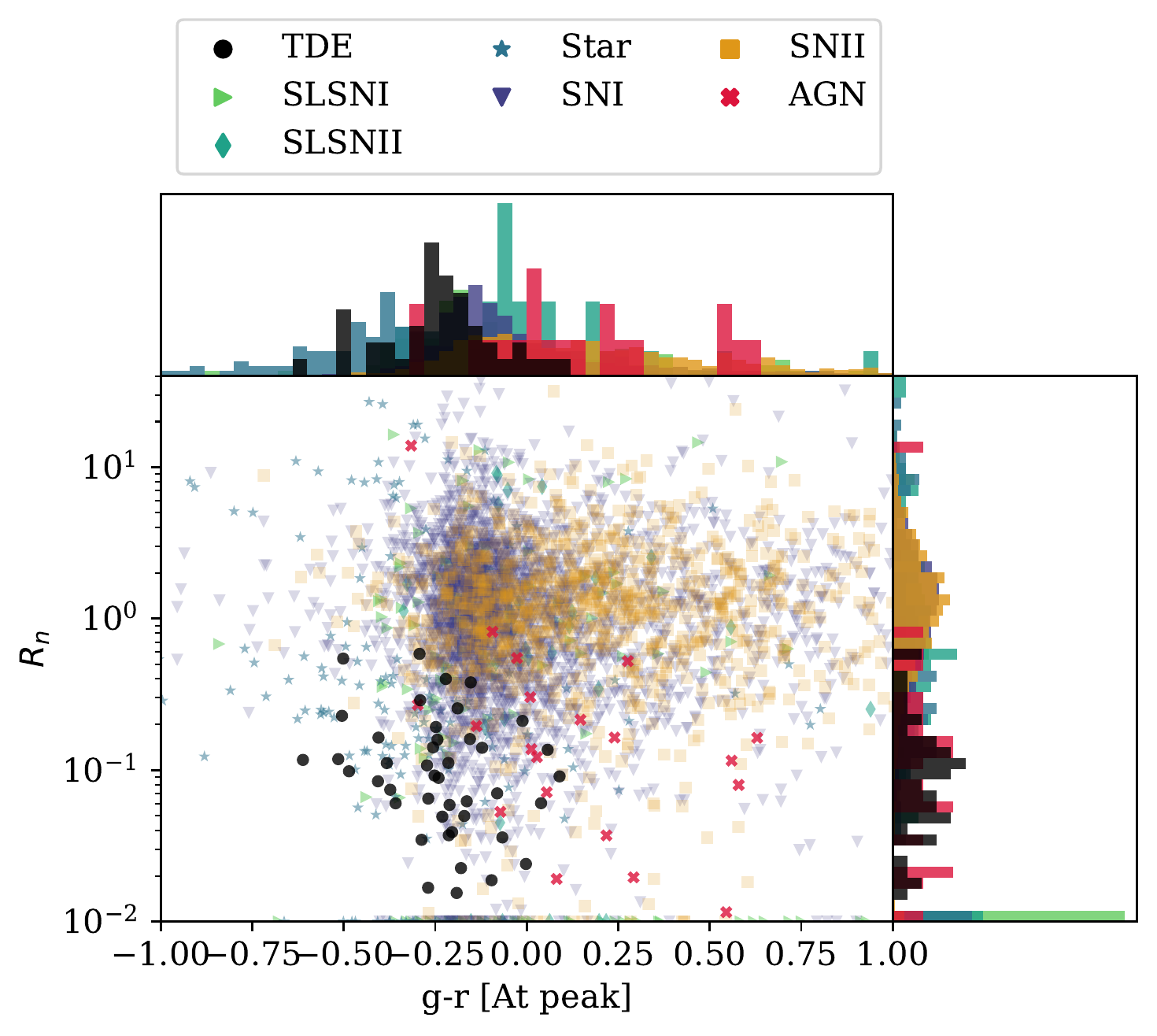}
\includegraphics[width=\columnwidth]{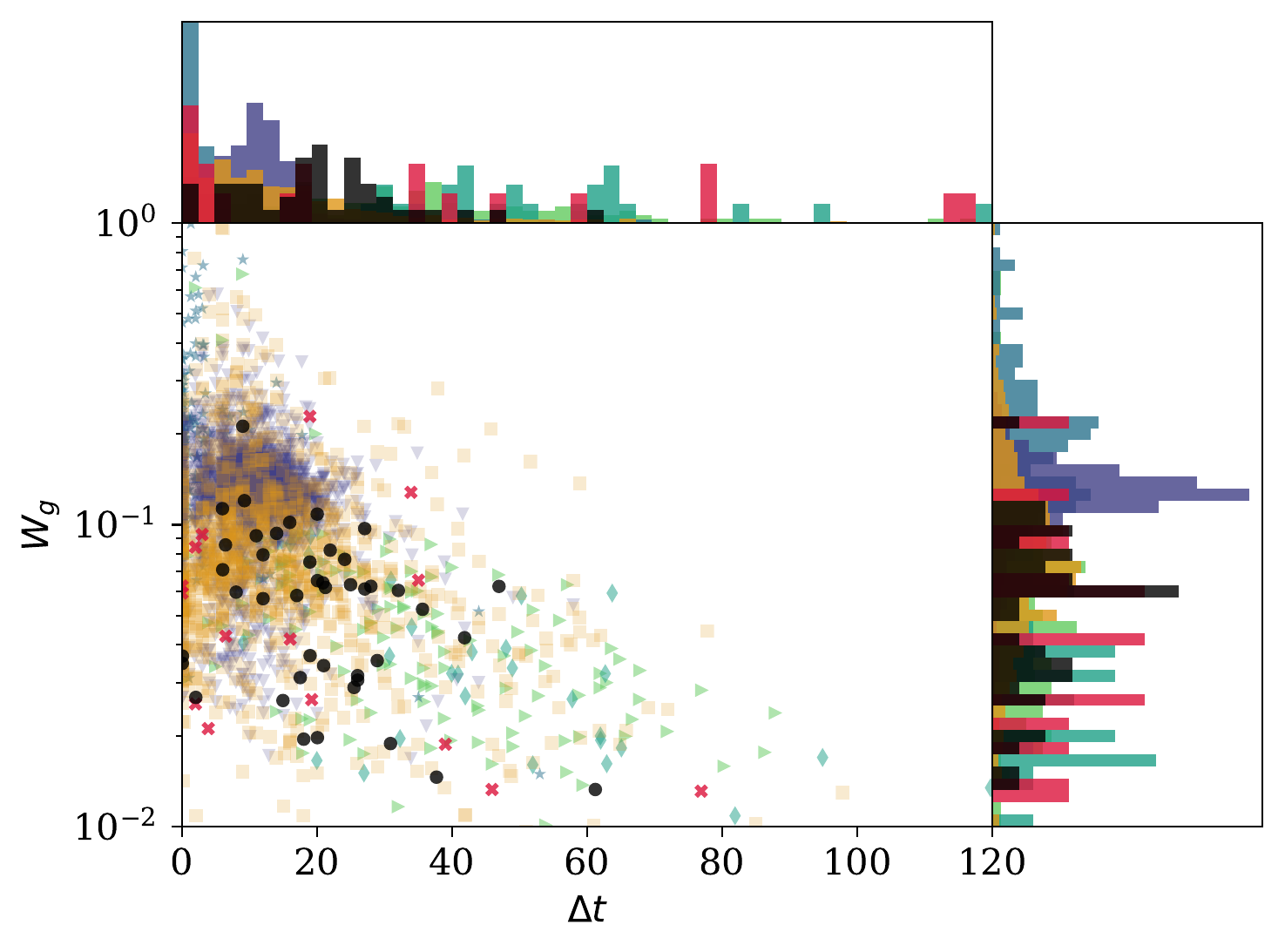}
\includegraphics[width=\columnwidth]{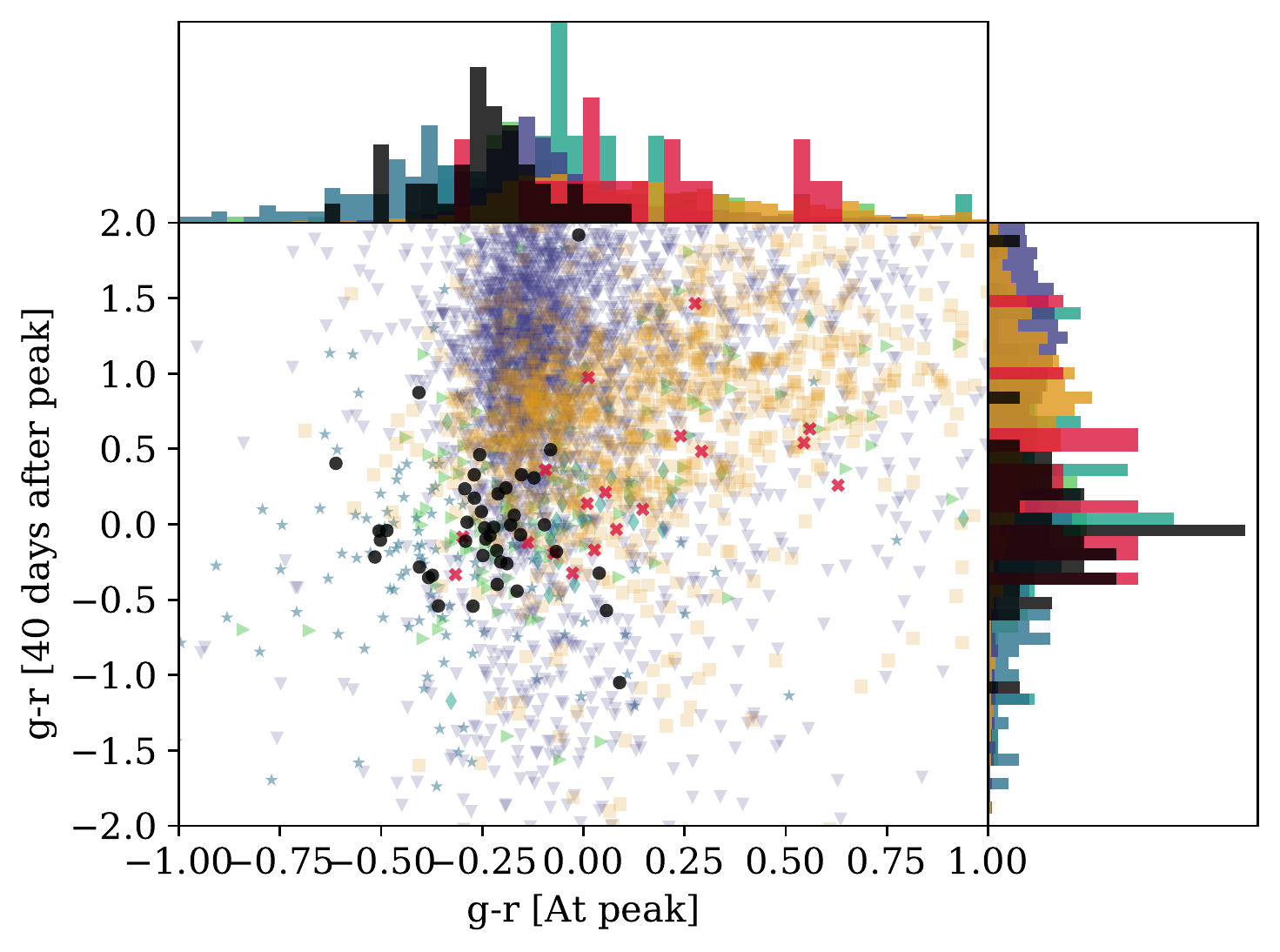}}
\caption{Phase space of the most relevant features used for the classifier, showing the various classes of transients in different colors/markers. \textit{Top}: The normalized host separation ($R_n$) versus color of the transient during peak $(g-r)_P$. \textit{Middle} : Width of the light curve in $g$-band ($W_g$) compared to the time to peak ($\Delta t$). \textit{Bottom} : color of the transient during peak $(g-r)_P$ compared to the $(g-r)_L$ color 40 days after peak. \label{fig:histogram}}
\end{center}
\end{figure}

FLEET is able to classify thousands of transients within a few hours on a personal computer. Classifying a new transient takes on the order of $10-20$ seconds. Once the required catalog data and an image of the host galaxy have been downloaded ($\sim 100$~KB each), the required time to re-classify a transient is about $5 - 10$ seconds.

\subsection{Feature Optimization}\label{sec:features}

We optimize several hyper-parameters of the RF classifier, including the number of days of photometry to consider, the features to be included in each model, and the depth of the RF trees. The two metrics we optimize for are the ``Completeness" and ``Purity". Completeness is defined as the total number of true positive TDEs divided by the total number of TDEs, and purity is the total number of true positive TDEs divided by the sum of true positive TDEs and false positive TDEs.

Our predictive accuracy improves as a transient evolves and more light curve data are used, but at the expense of identifying the transient at a later phase. We optimize the number of days of photometry to include in both the rapid and late-time classifiers, testing the performance of FLEET using 10 to 80 days of photometry in steps of 5 days (Figure~\ref{fig:days_tde}). We find that for the rapid classifier, including 20 days of photometry is the minimum required to produce reasonable results, since including only 15 days of photometry reduces the purity and completeness by about half. Conversely, we find that including 40 days of photometry in the late-time classifier significantly increases the completeness from $\approx 35$\% to $\approx 50$\% compared to using only 35 days. These thresholds are reasonable when we compare to the typical rise times of the transients in our training set, 95\% of which peak $\lesssim 30$ days after discovery. We find that including more than 40 days of photometry does not significantly improve either the purity or completeness. Including more than $\sim 70$ days of photometry actually begins to degrade our metrics, since these late-time data might be of lower quality, with higher scatter, or include phenomena such as a flattening or light curve bumps that are not accounted for in our model.

\begin{figure*}
    \begin{center}
    \centering
    {\includegraphics[width=\columnwidth]{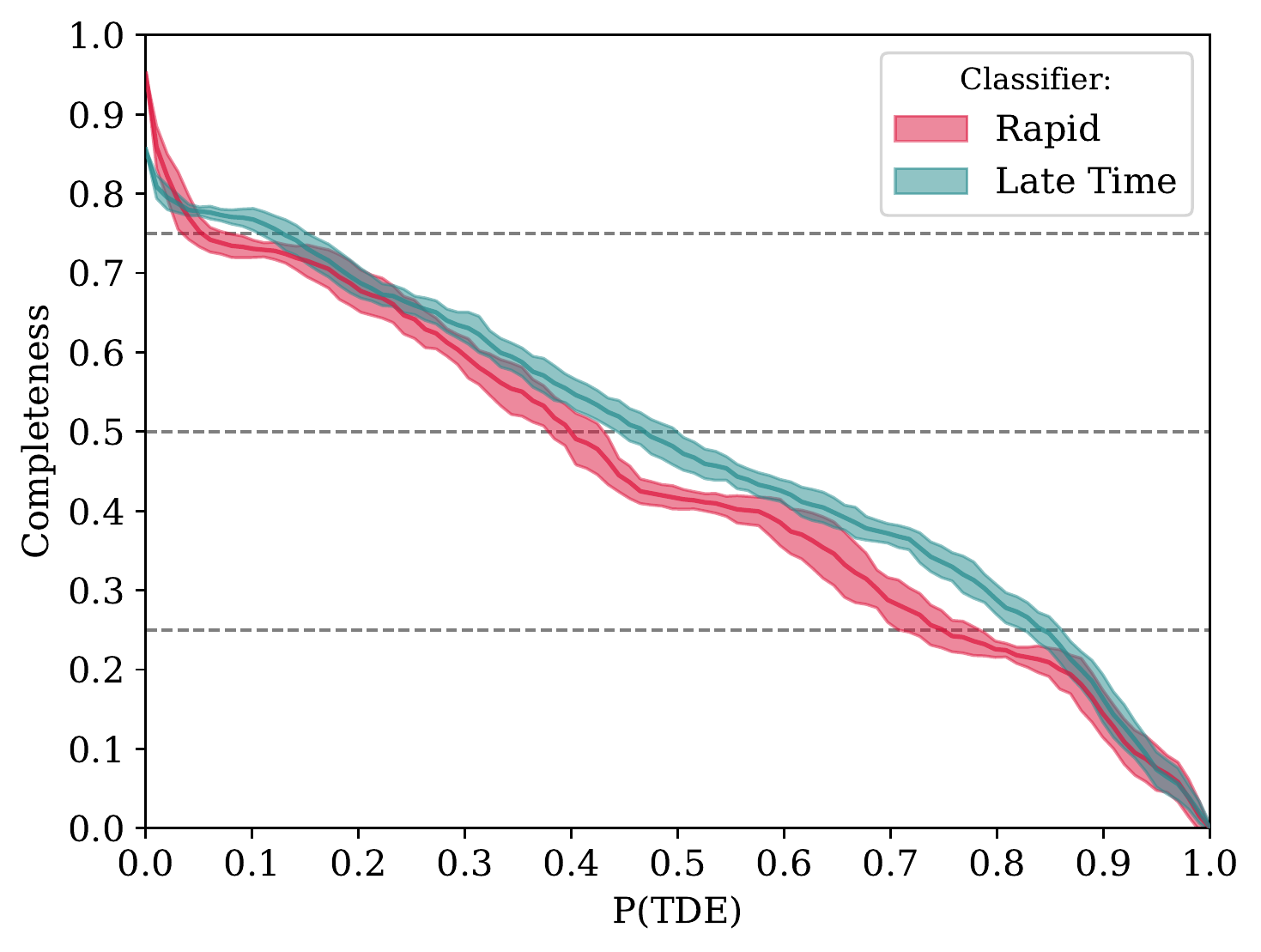}}
    {\includegraphics[width=\columnwidth]{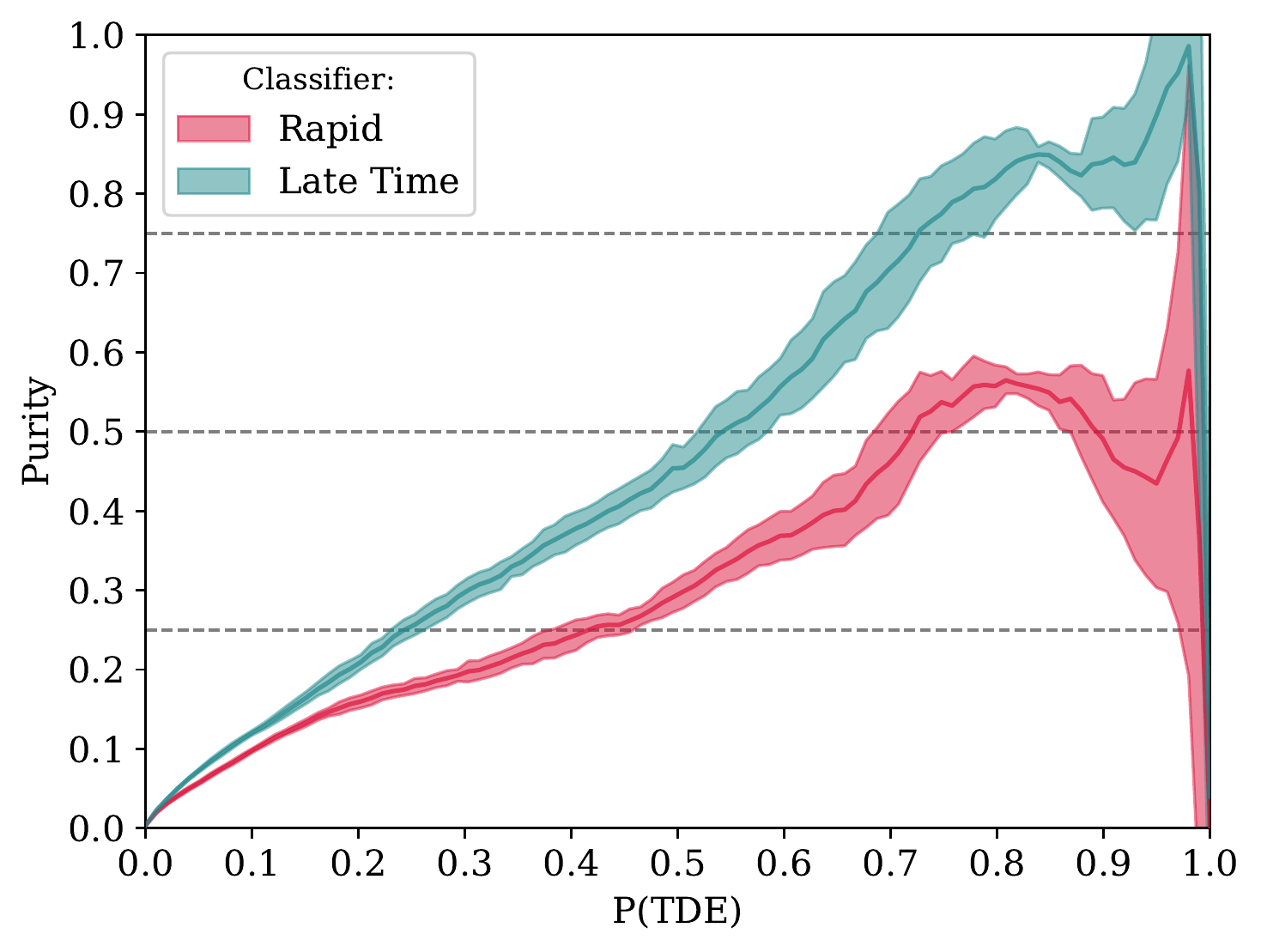}}
    \caption{\textit{Left}: Completeness as a function of classification confidence for the rapid and late-time classifiers. \textit{Right}: Purity as a function of classification confidence for the rapid and late-time classifiers. The shaded regions represent $1\sigma$ uncertainties, calculated by generating each model using 25 different random seeds. \label{fig:all_features} }
    \end{center}
\end{figure*}

Additionally, we optimize the set of features to include in the classifier and determine the optimal set that yields the highest purity to be:

\begin{itemize}
    \item $W$ : the width parameter of the light curve.
    \item $A$ : multiplicative modifier for the decline rate of the light curve (only used for late-time model).
    \item $R_n$: transient-host separation normalized by half-light radius of the host.
    \item $\Delta t$: the time of peak brightness minus the time of discovery.
    \item $\Delta m$: host magnitude minus the peak observed transient magnitude.
    \item $(g-r)_P$: the light curve color at peak.
    \item $(g-r)_L$: the late-time light curve color, 40 days after peak.
\end{itemize}

We use one $W$, $A$, and $\Delta m$ for each, $g$- and $r$-band. The half-light radius in $r$-band of the host galaxy is obtained from the SDSS catalog, or from the PS1/$3\pi$ catalog if the object is not in SDSS. The $g-r$ color is measured from the light curve fits, as opposed to the photometry, since we found this to be a more uniform method that allows us to measure an estimate for the color even for transients with sparse data coverage. During our optimization process, we tested the effects of including other features in the classifier, including: the photometric redshift of the host galaxy, the $g-r$ color measured from the brightest observed photometry data points, the host separation, the host radius, a $\chi^2$ estimate for goodness of fit of the model, and the probability of chance coincidence $P_{\rm cc}$. We found that including these features either hurt or did not improve our metrics.

The importance of each included feature is not defined independently of other features; if two features are correlated then their relative importance could be affected. We use the permutation importance method described in \cite{Breiman01} to calculate the correlated importance of each feature, and show the results for the late-time classifier in the top panel of Figure~\ref{fig:correlation}. In the bottom panel of Figure~\ref{fig:correlation} we show the correlation between features and find that with the exception of a strong correlation between $\Delta m_g$ and $\Delta m_r$, the individual features are mostly independent. We find the most important features to be $R_n$, and $(g-r)_P$ for both the rapid and late-time classifiers, and also $(g-r)_L$ for the late-time classifier.

In Figure~\ref{fig:histogram} we plot the phase-space of our most important features that distinguish TDEs from the other transients used for training. The figures show how $R_n$ and $(g-r)_P$ can help separate TDEs from other transients. A low $R_n$ value is obviously a good discriminant since TDEs happen in the nuclei of galaxies. $(g-r)_P$ can help separate TDEs from AGN and SNe, since TDEs tend to be bluer. Lastly, $(g-r)_L$ becomes important when selecting TDEs, since their colors tend to remain fairly constant as they evolve, unlike other transients (e.g., \citealt{Velzen20, Hammerstein22, Nicholl22}).

Finally, we fit for the optimal depth of the RF trees. We optimize the tree depth by running a grid search from a depth of $5$ to $20$ in steps of three, then selecting the three values with the highest purity and repeating the same test using a finer grid with steps of 1. We find an optimal depth of $14$ for the rapid classifier and $17$ for the late-time classifier.

\subsection{Validation}

\begin{figure*}
    \begin{center}
    \centering
    {\includegraphics[width=0.8\columnwidth]{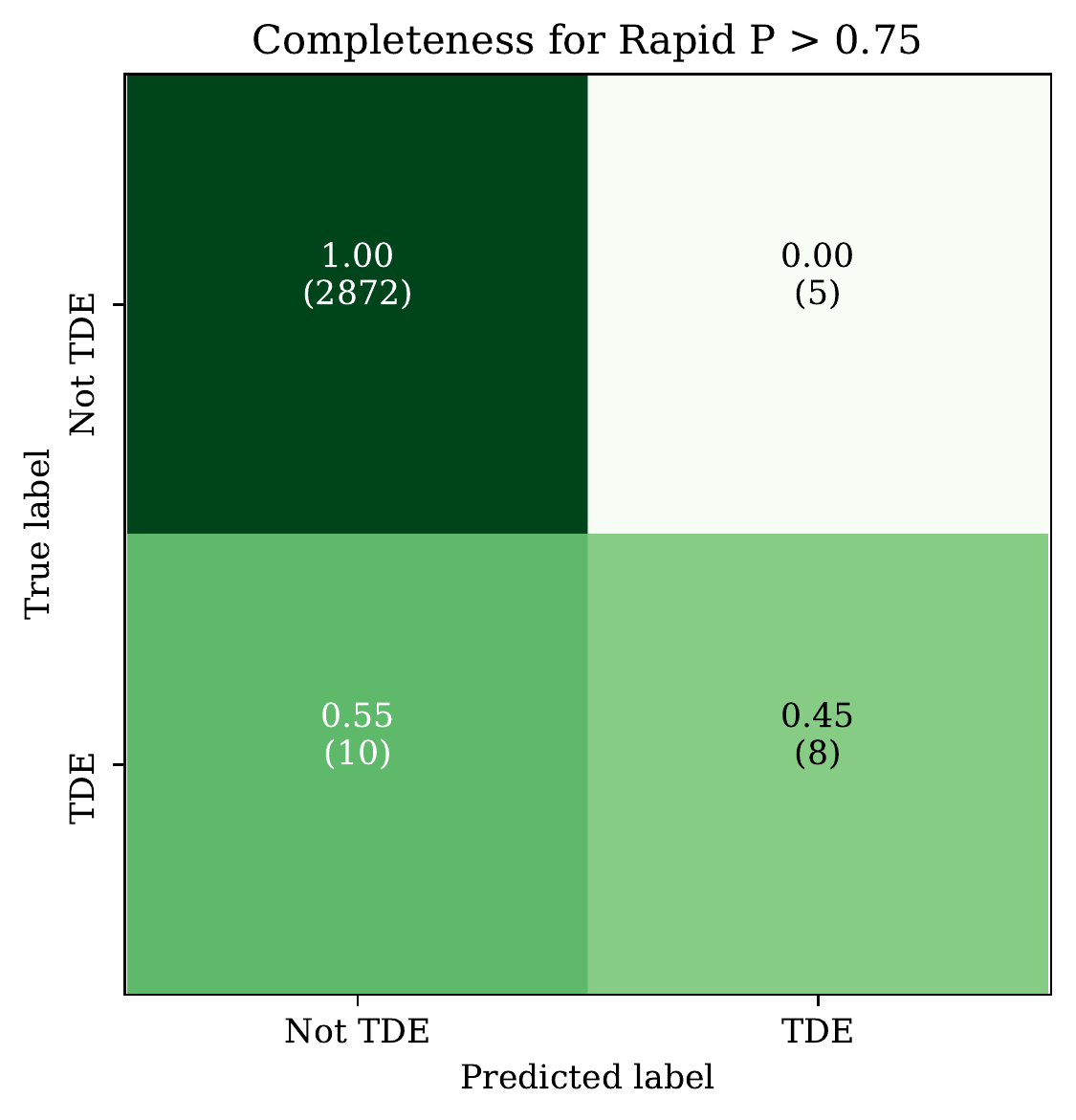}
    \includegraphics[width=0.8\columnwidth]{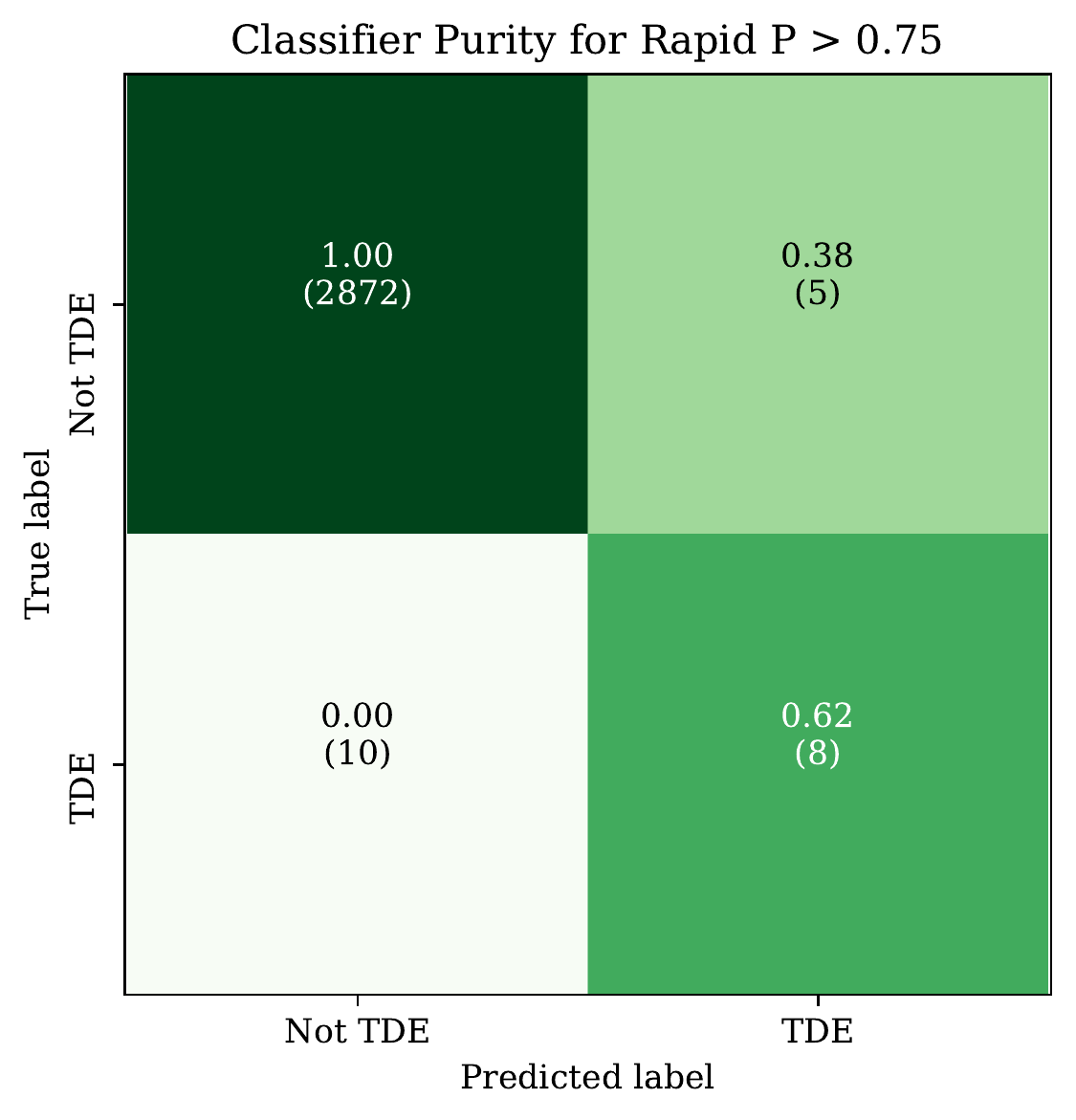}
    \includegraphics[width=0.8\columnwidth]{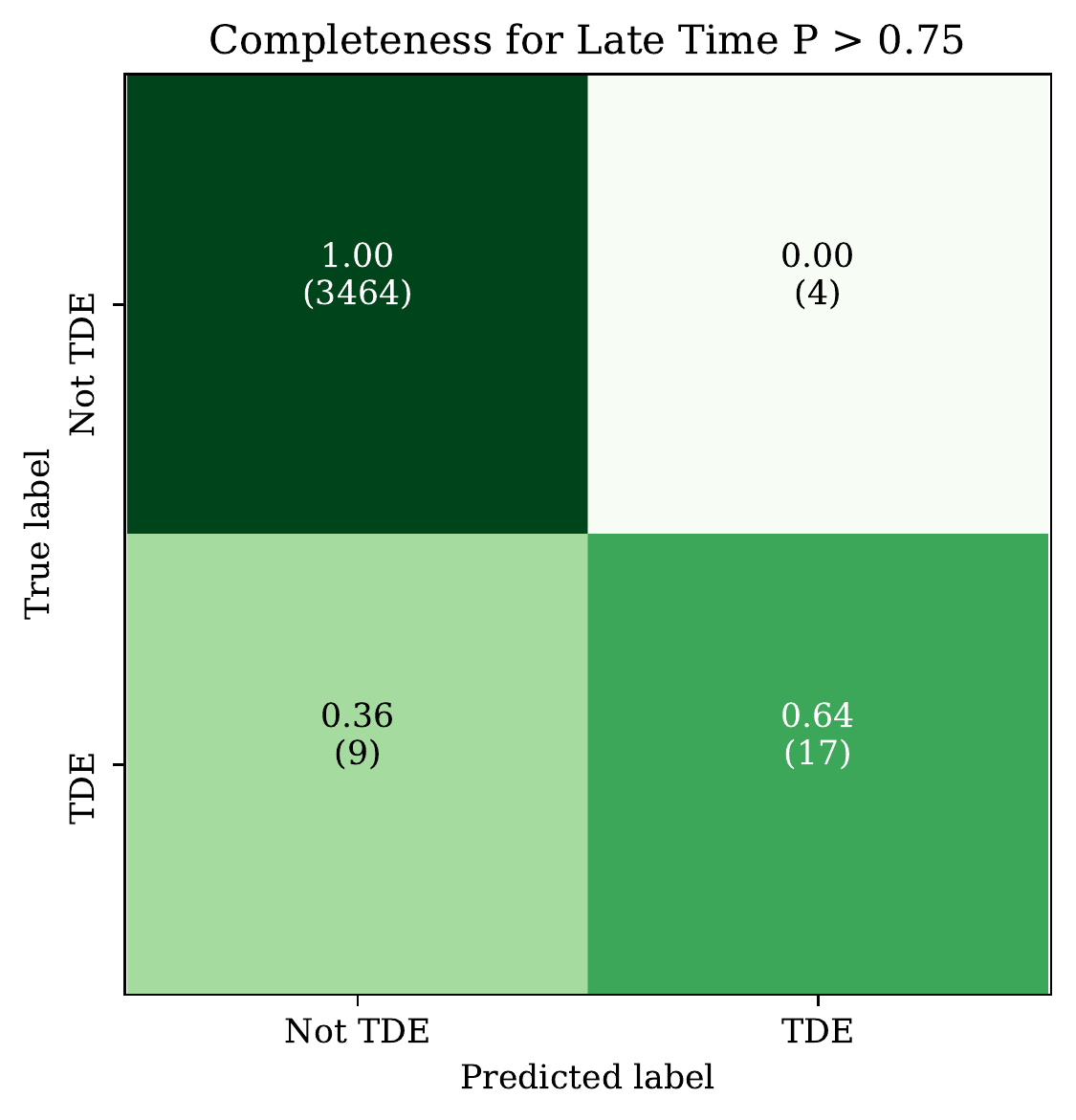}
    \includegraphics[width=0.8\columnwidth]{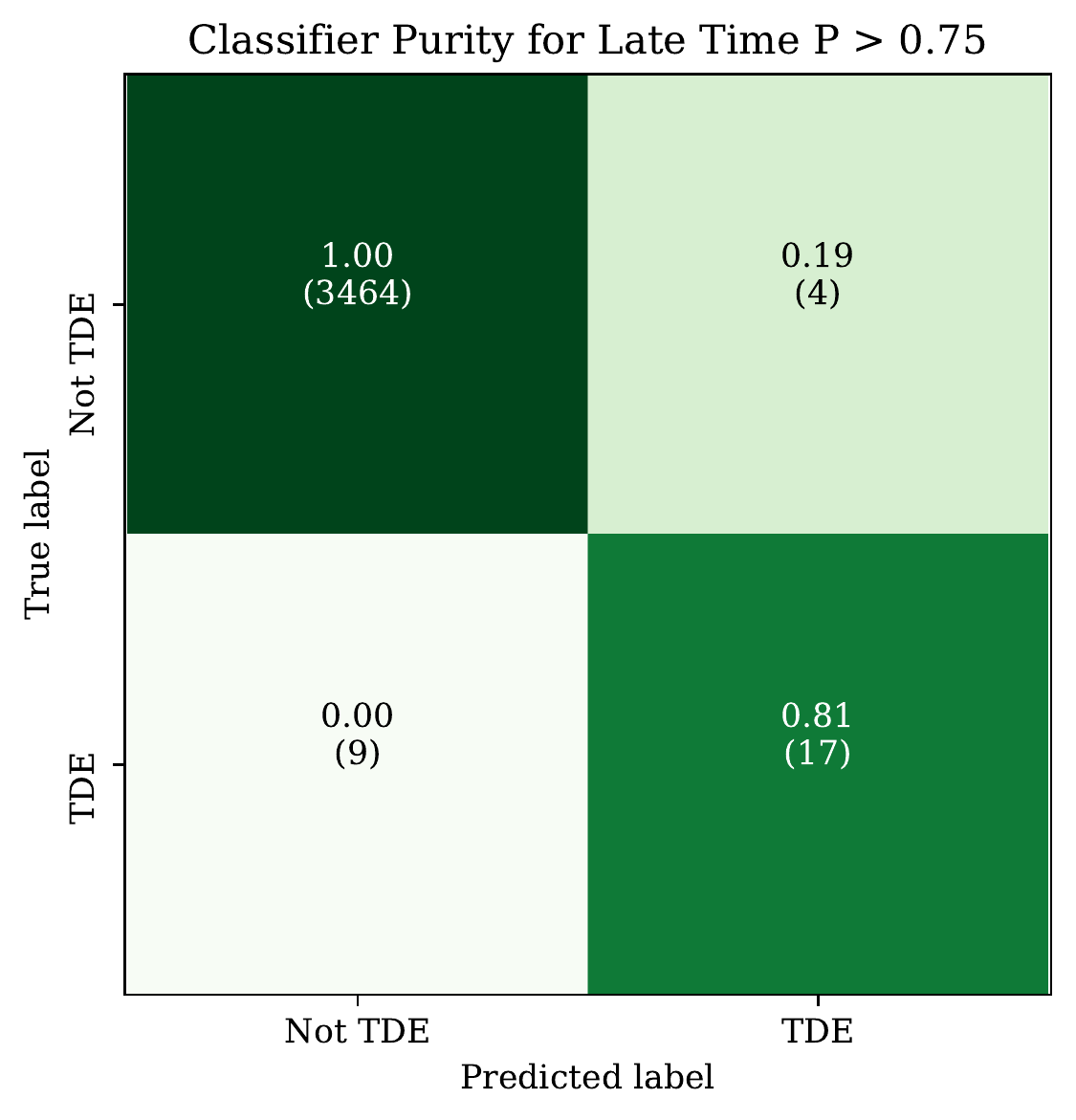}}
    \caption{Using a sample of only transients with a classification probability of \ptde$> 0.75$ or $P$(not-TDEs) $> 0.75$, we produce a confusion matrix that indicates a purity of 62\% for the rapid classifier and 81\% for the late-time classifier. The corresponding completeness values for the same classifiers are 45\% and 64\%. These values represent the raw number of TDEs used in the classifier and are not weighted by the number of samples in each class. \label{fig:confusion}}
    \end{center}
\end{figure*}

For validation of the algorithm, we implement a leave-one-out cross-validation method. This method trains the classifier using every transient except one, predicts the classification of the one transient, and then repeats this process cycling through all transients. This allows us to robustly test our classifier without having to divide the data set into a training and test set, which would compromise the already small sample of TDEs. We use three different methods to evaluate the performance of our classifier: a confusion matrix, a purity curve, and a completeness curve. Since we are not concerned with the classification of transients other than TDEs, we collapse the individual transient classification categories listed in \S\ref{sec:fleet} into a binary ``TDE" versus ``non-TDE" classification. To calculate the non-TDE probability for each transient we sum the probabilities of all other transient classes.

The purity of the classifier increases and the completeness declines as we restrict the sample to events with progressively higher values of classification confidence (Figure~\ref{fig:all_features}). For \ptde$>0.5$, the purity is $\approx 45\%$ for the late-time classifier and $\approx 30\%$ for the rapid classifier, at a corresponding completeness of $\approx 50\%$ and $\approx 40\%$, respectively. This represents a factor of $\sim 60$ improvement over a random selection of transients, which would yield a  $\approx 0.5\%$ success rate in a magnitude-limited survey \citep{Fremling20,Villar19}.

In Figure~\ref{fig:confusion} we show the confusion matrices, namely, the label predicted by our classifier compared to the true label of the transient, normalized both in terms of purity and completeness. We impose a confidence cut of $P>0.75$ for either the TDE or not-TDE classes, corresponding to the peak classifier purity. We find that 64\% ($N = 17$) and 45\% ($N = 8$) of true TDEs were predicted to be TDEs by the late-time and rapid classifiers, respectively. The matrices also show that 81\% ($N = 17$) and 62\% ($N = 8$) of all transients predicted to be TDEs were true TDEs, for the late-time and rapid classifiers, respectively.

\begin{figure}
    \begin{center}
    \centering
    {\includegraphics[width=\columnwidth]{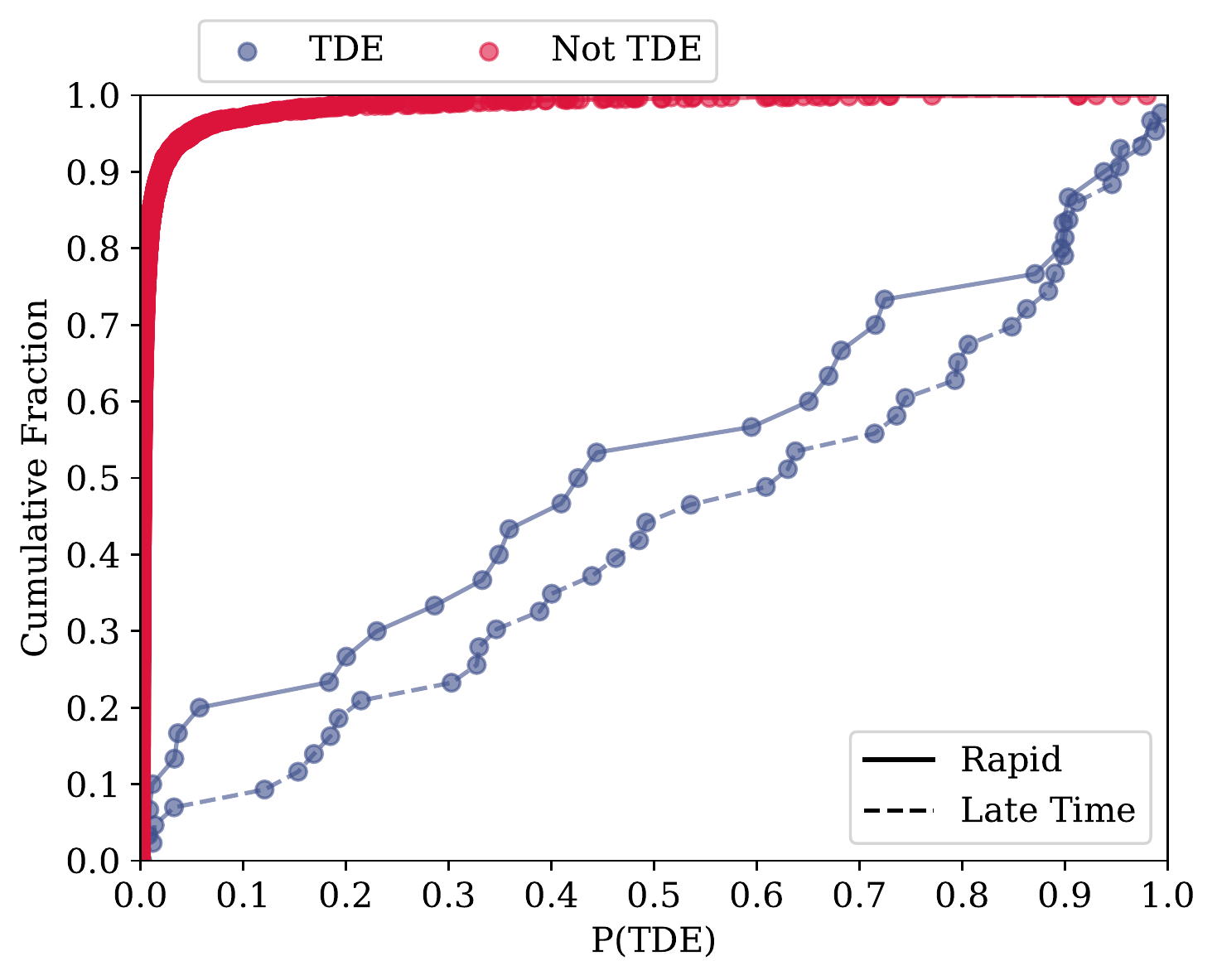}}
    \caption{Cumulative distribution as a function of classification confidence \ptde\ for transients that are true TDEs (blue), or not TDEs (red). We find that most non-TDEs are correctly identified as such, while half of true TDEs have a \ptde$\gtrsim$0.6. The rapid and late-time classifier lines for non-TDEs appear indistinguishable from each other in the plot.\label{fig:crossy}}
    \end{center}
\end{figure}

In Figure~\ref{fig:crossy} we show how the rapid and late-time classifiers perform at classifying TDEs and not misclassifying other objects, as a function of classification confidence level, \ptde. We find that half of the correctly identified TDEs have a \ptde$\gtrsim 0.6$, but some true TDEs are still misclassified with a low \ptde$\lesssim 0.1$. Some of the misclassified TDEs include AT\,2018jbv, AT\,2020ddv, AT\,2020opy, and AT\,2020riz, which were instead classified as likely SLSNe due to their relatively high $\Delta m$; and AT\,2019gte and AT\,2020neh, which were classified as likely SNe Ia due to their relatively fast light curves and lack of late-time photometry.

As an additional point of comparison, we test how the performance of FLEET compares to the performance of a more classical approach of targeting TDEs by imposing basic selection cuts on their observational parameters (e.g., \citealt{Velzen20}). We test the effectiveness of using selection cuts on the two most important parameters, $(g-r)_P$ and $R_n$. We find that imposing a cut and only selecting targets with $(g-r)_P < -0.5$ mag yields the highest purity of $\sim 5$\% at a corresponding completeness of $\sim 10$\%. Alternatively, imposing a cut of $(g-r)_P < 0$ mag yields a much higher completeness of $\sim 85$\%, but with an even lower purity of $\sim 3$\%. Similarly, we determine $R_n < 0.14$ to be the optimal threshold that yields the highest purity of $\sim 7$\% at a corresponding completeness of $\sim 60$\%. The purity obtained from either a cut on $(g-r)_P$ or $R_n$ is well below the $\approx 50$\% purity we can obtain from FLEET, even with the rapid classifier. We further attempt to optimize the purity of these selection cuts by implementing them in conjunction. We find that a combined threshold of $(g-r)_P < -0.5$ mag and $R_n < 0.12$ would yield the highest possible purity of $\sim 38$\%. This purity is much closer to the estimates from FLEET, but comes at the expense of a $\sim 6$\% completeness, much lower than the $\approx 50$\% completeness that FLEET can achieve at the same purity level. Therefore, we have shown that FLEET is expected to perform much better than imposing basic selection cuts.

\subsection{Classifier Summary}

Using our rapid classifier, trained using host galaxy information and the first 20 days of photometry of 45 TDEs, we predict we can recover TDEs with a $\approx 50$\% purity at $\approx 30$\% completeness for \ptde$>0.7$. With our late-time classifier, trained instead on 40 days of photometry, we can recover TDEs with a purity of $\approx 70$\% and a corresponding completeness of $\approx 40$\% for the same \ptde$>0.7$ threshold.

Currently, time-domain surveys report $\sim 20,000$ transients per year, and we expect $\sim 0.5$\% of them to be TDEs given current observational rates, corresponding to $\sim 100$ TDEs that could be discovered every year. Given our predictions for purity and completeness, using FLEET would result in $\sim 15 - 28$ TDEs discovered per year, depending on the classifier used. This is about twice as many as the $8 - 16$ TDEs classified each year since 2019.

\section{Selection Effects}\label{sec:selection}

\begin{figure*}
    \begin{center}
    \centering
    {\includegraphics[width=0.9\textwidth]{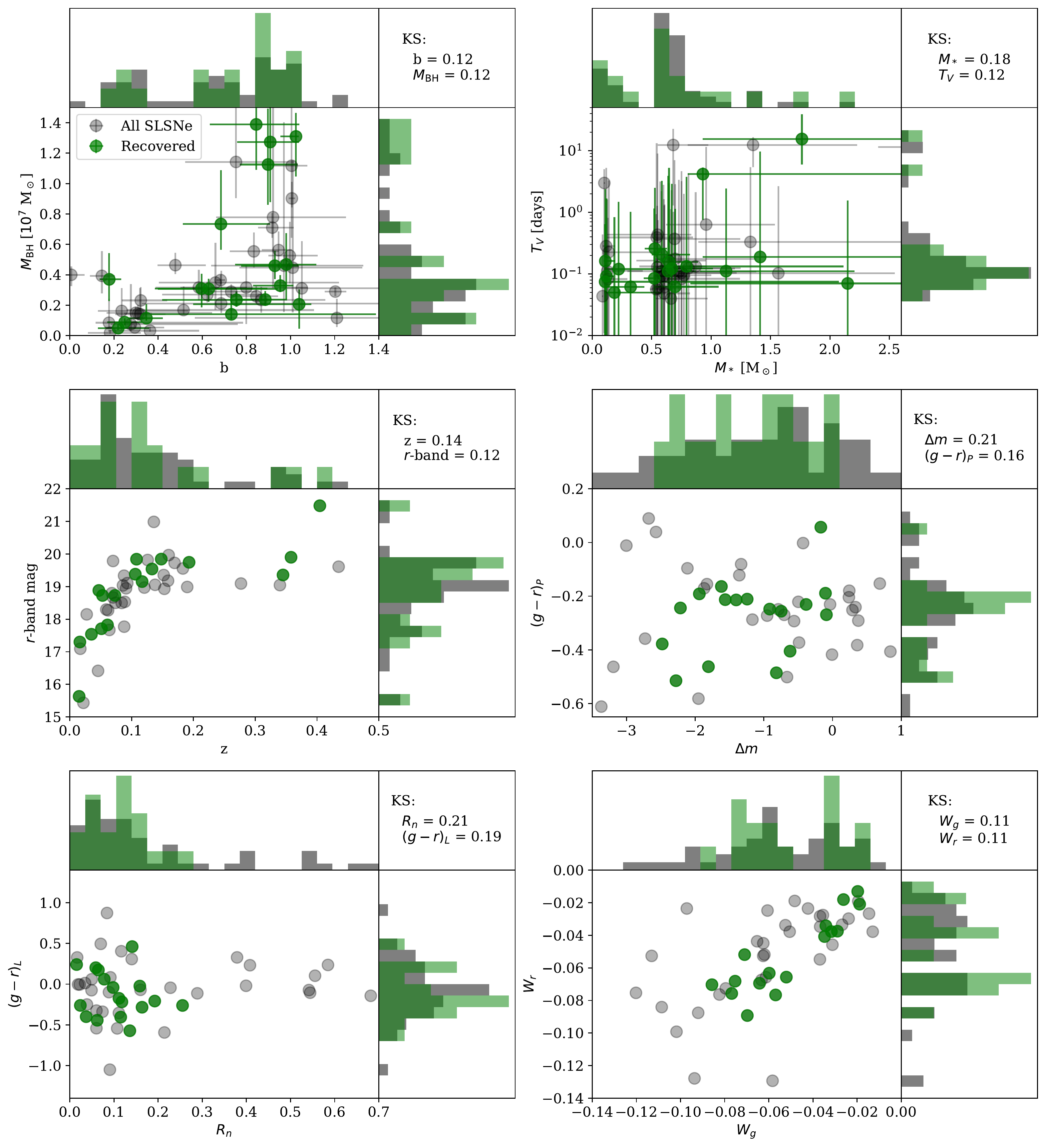}}
    \caption{The gray data points show the physical and observational parameters from a sample of TDEs obtained from \cite{Nicholl22}. The green data points are TDEs that were recovered by FLEET with a \ptde$> 0.5$. \textit{Top Left}: Black hole mass and impact parameter. \textit{Top Right}: stellar mass and viscous timescale. \textit{Middle Left}: redshift and apparent peak $r$-band magnitude. \textit{Middle Right}: $\Delta m$ compared to the $(g-r)_P$ color at peak. \textit{Bottom Left}: late-time color $(g-r)_L$ compared to the normalized host separation $R_n$. \textit{Bottom Right}: The width of the light curve in $g$ and $r$ bands. We include the KS metric for each parameter on the upper right corner of each sub-plot. We find no obvious bias against physical or observational parameters that are not already features of the classifier. \label{fig:bias} }
    \end{center}
\end{figure*}

We explore possible selection effects that might result from using FLEET for target selection. For testing, we classify the 32 TDEs presented in \cite{Nicholl22} using the late-time FLEET classifier and compare the observational parameters of the TDEs with \ptde$>0.5$ with those of all TDEs. To quantify if the parameter distributions are different, we implement a two-sample Kolmogorov-Smirnov (KS), where a KS metric of $D = 0$ indicates the two samples are drawn from the same distribution, and $D = 1$ means there is no overlap between the distributions. In Figure~\ref{fig:bias} we show how the two populations compare and find no obvious bias against most parameters. We determine KS metrics (and $p$ values of) $0.14 (0.93)$ for redshift, $0.12 (0.98)$ for apparent magnitude, $0.21 (0.52)$ for $R_n$, $0.21 (0.52)$ for $\Delta m$, $0.16 (0.85)$ for $(g-r)_P$, $0.11 (0.99)$ for $W_g$ and $W_r$, and $0.19 (0.64)$ for $(g-r)_L$, meaning the two populations differ at the $\sim 10-20$\% level.

Most of the parameters shown in Figure~\ref{fig:bias} show little difference between the full population of TDEs and the population of recovered TDEs with \ptde$>0.5$. The biggest difference is for $R_n$, where TDEs with $R_n \gtrsim 0.3$ do not tend to be selected by FLEET. This is not to say that transients with $R_n > 0.3$ are not nuclear, but simply that the uncertainty in their coordinates is high, but still consistent with being nuclear. This uncertainty is dominated by the typical scatter in ZTF coordinates, which can be up to $\sim 0.5''$. We find that TDEs with $\Delta m \gtrsim 0$ or $W_g \lesssim -0.1$ are also less likely to be selected as TDEs.

Similarly, we explore the difference in physical parameters for the recovered TDEs compared to the full TDE sample. For the exploration of physical parameters, we use the TDE models from \cite{Nicholl22}, who fit the light curves of 32 TDEs using {\tt MOSFiT} \citep{guillochon18}. We explore the differences in the impact parameter $b$, black hole mass $M_{\rm BH}$, stellar mass $M_*$, and viscous timescale $T_V$, and find no obvious selection effects. We find a KS metric (and $p$ value) of $0.12 (0.94)$ for $b$, $0.12 (0.99)$ for $M_{\rm BH}$, $0.18 (0.67)$ for $M_*$, and $0.12 (0.97)$ for $T_V$, indicating that the parameter distributions for the full TDE sample and the sample of recovered TDEs with \ptde$>0.5$ are very similar and only differ at the $\sim 10$\% level.

In conclusion, we find that even if FLEET selects events with certain observational properties, these do not translate to selection effects in physical parameters. In other words, FLEET does not appear to produce biases against physical or observational parameters not included as features in the classifier. Of course this is in relation to the known sample of TDEs; finding anomalous TDEs not found in the current known sample, such as off-nuclear TDEs or white dwarf TDEs, will likely require larger samples from systematic surveys such as \textit{Rubin} or \textit{Roman}.

\startlongtable
\begin{deluxetable*}{cc|cc|cc|cc}
    \tablecaption{Likely TDEs from ZTF \label{tab:ztf}}
    \tablehead{\colhead{Name}   & \colhead{\ptde}      &   \colhead{Name} & \colhead{\ptde} &   \colhead{Name} & \colhead{\ptde}  &   \colhead{Name} & \colhead{\ptde}  }
    \startdata
        2021crk  & 0.97 & 2021ldl & 0.80 & 2020kri  & 0.67 & 2020aexc & 0.58 \\
        2019gtm  & 0.97 & 2020ygl & 0.80 & 2021pqg  & 0.66 & 2021qbh  & 0.57 \\
        2021aees & 0.91 & 2019zbt & 0.80 & 2018jil  & 0.66 & 2020afap & 0.57 \\
        2021aeuf & 0.90 & 2021zvy & 0.72 & 2021uhs  & 0.64 & 2019aamf & 0.57 \\
        2021lsi  & 0.86 & 2021ony & 0.71 & 2021xbz  & 0.63 & 2021abkx & 0.56 \\
        2020bgf  & 0.86 & 2020qfm & 0.71 & 2021acqt & 0.63 & 2019enj  & 0.56 \\
        2021kqp  & 0.85 & 2019phf & 0.71 & 2019aami & 0.62 & 2021aux  & 0.52 \\
        2021wdh  & 0.82 & 2021icq & 0.70 & 2021nhq  & 0.61 & 2020qmx  & 0.51 \\
        2021iqs  & 0.82 & 2019qdh & 0.70 & 2019saj  & 0.60 & 2020aawn & 0.50 \\
        2020dxw  & 0.81 & 2019wlv & 0.69 & 2020nmc  & 0.58 &          &      \\
    \enddata
    \tablecomments{List of 39 unclassified ZTF transients with a high likelihood of being a TDE, ordered by \ptde.}
\end{deluxetable*}

\section{Implementation of FLEET For Time-Domain Surveys}\label{sec:surveys}

\subsection{ZTF}\label{sec:ztf}

ZTF is the survey that currently reports the most transients to the TNS, but $\sim 90$\% of them remain spectroscopically unclassified. Here, we use the late-time FLEET classifier to obtain a \ptde\ estimate for all ZTF transients reported to the TNS with the aim of recovering TDEs that were previously missed.

At the time of writing, there were a total of 95,729 transients reported to the TNS. Since we are using the late-time classifier, we exclude transients that are too young, discovered after 1 March 2022. Additionally, FLEET only works for transients within the PS1/$3\pi$ footprint, and with at least 2 $g$-band and 2 $r$-band points. After applying these cuts we run FLEET on a final list of 31,892 transients, and list 39 likely TDE candidates with \ptde$\geq0.5$ obtained from this experiment in Table~\ref{tab:ztf}. Follow-up and analysis of these targets will be presented in future work.

Given their individual probabilities of being TDEs, if we were to spectroscopically classify these 39 TDE candidates we would expect $\sim 28$ of them to be confirmed as TDEs, which would represent a factor of $\sim 50$\% increase in the current population of TDEs. These tools will be more powerful for larger data sets like the ones \textit{Roman} or \textit{Rubin} will produce, allowing us to generate large photometrically selected samples of TDEs and analyze their statistical demographics without the need for spectra.

\begin{figure}
    \begin{center}
    \centering
    {\includegraphics[width=\columnwidth]{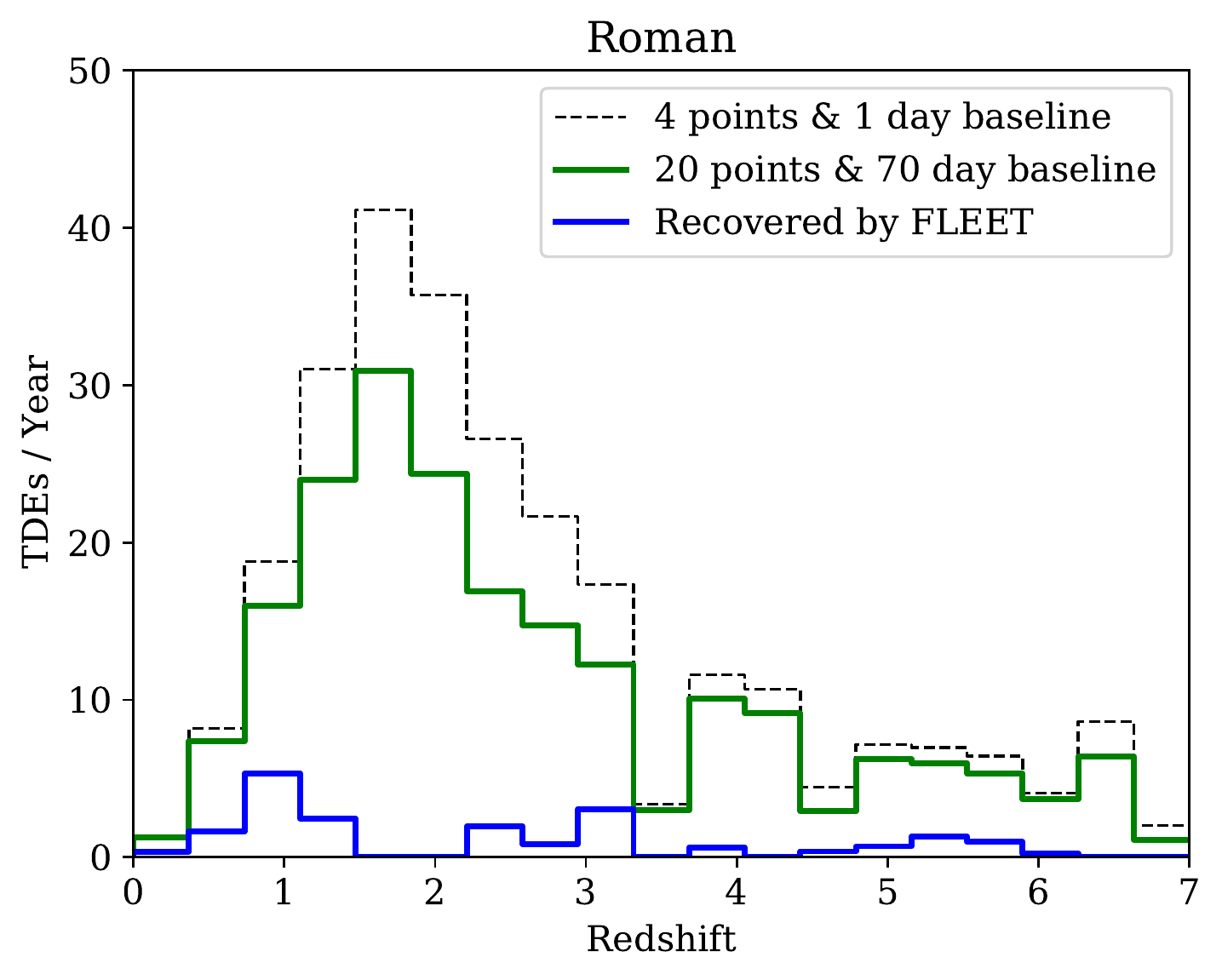}}
    \caption{Number of TDEs expected to be found by the \textit{Roman} HLTDS per year, per redshift bin. The black dashed line shows the total number of TDEs with at least 4 detections spanning at least a 1 day baseline, while the green solid line shows only the ``well observed" TDEs with at least 20 detections over a 70 day baseline. The blue line shows the TDEs recovered by FLEET with \ptde$>0.5$. \label{fig:counts_Roman}}
    \end{center}
\end{figure}

\begin{figure*}
    \begin{center}
    \centering
    {\includegraphics[width=0.9\textwidth]{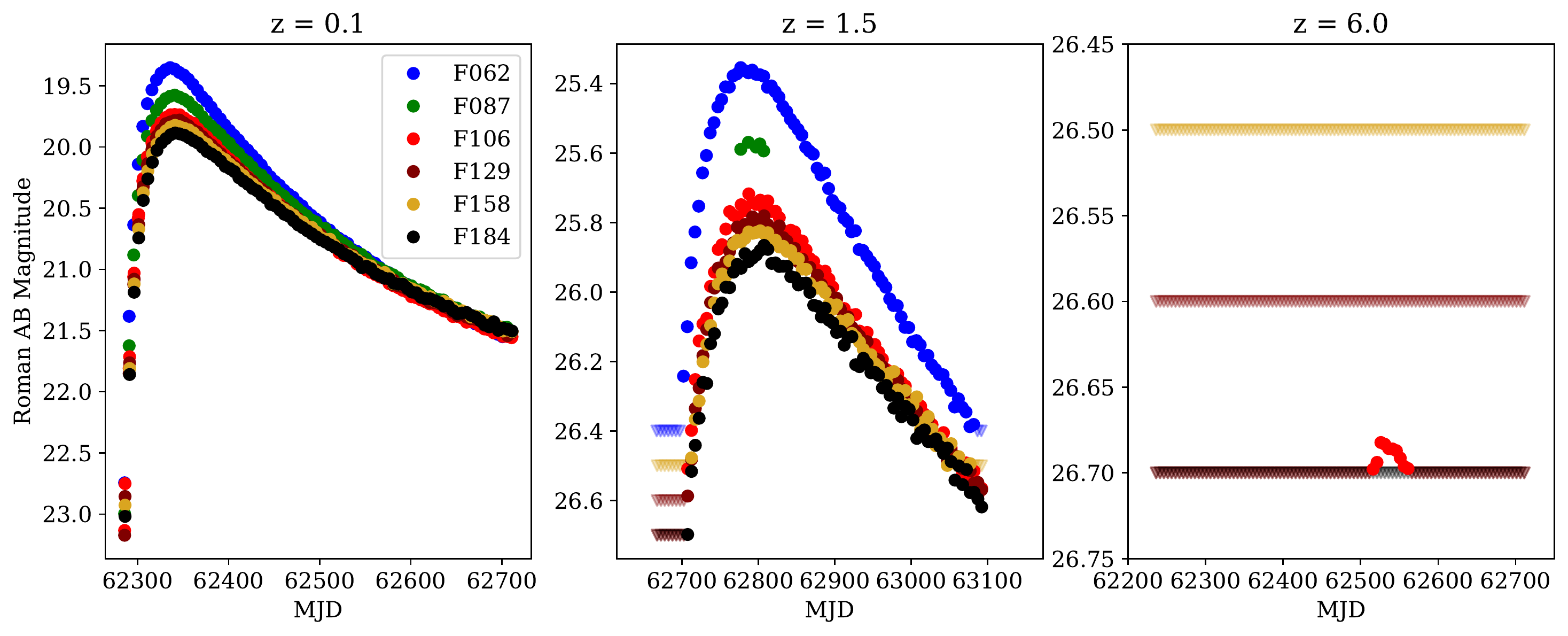}}
    {\includegraphics[width=0.9\textwidth]{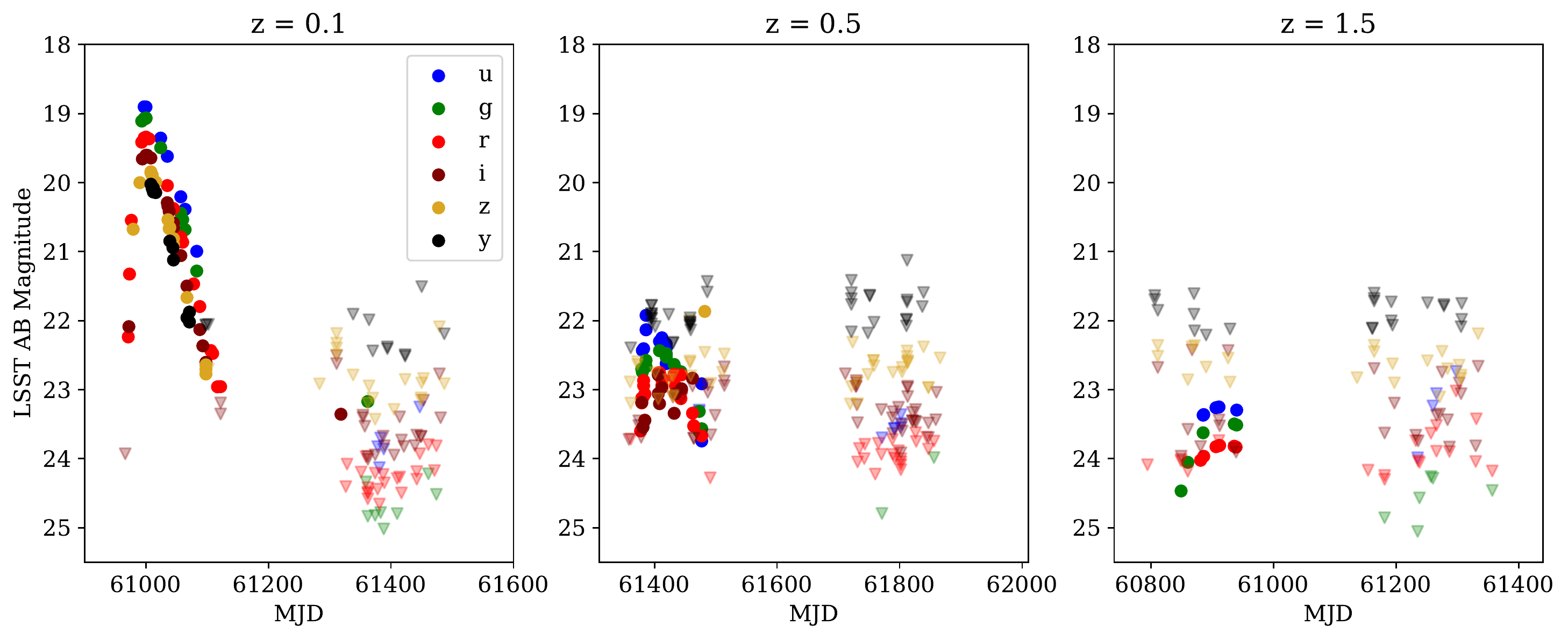}}
    \caption{Representative model light curves of TDEs from the \textit{Roman} HLTDS (\textit{Top}), and \textit{Rubin} (\textit{Bottom}) simulated surveys. We show three light curves for each survey, at three different redshifts. \label{fig:lightcurves} }
    \end{center}
\end{figure*}

\subsection{Roman}\label{sec:roman}

The \textit{Nancy Grace Roman Space Telescope} (\textit{Roman}) is an infrared wide-field survey telescope planned to commence observations in 2027 \citep{Spergel15}. A major component of the \textit{Roman} mission are the \textit{community surveys}, large programs that will occupy a large fraction of the mission and are meant to provide data useful for the community at large. One such survey is the High Latitude Time Domain Survey (HLTDS), which focuses on the study of extragalactic transients \citep{Rose21}. Here, we estimate the number of TDEs that could be found by the \textit{Roman} HLTDS, and explore the possibility of using FLEET to find them.

The final survey parameters of the HLTDS have not been defined, but we simulate a \textit{Roman} HLTDS observation sequence based on the parameters suggested in \cite{Rose21}. We exclude the other two community surveys from our simulation since the Galactic Bulge Time Domain Survey will be heavily crowded by stars and affected by Galactic extinction, and the High Latitude Wide Area Survey has no time domain component. We run a simulation that consists of 146 visits of a patch of sky located at RA = 25 deg., DEC = $-53$ deg. with a cadence of 5 days. Each visit has a wide and narrow component. The wide component covers an area of 15.6 deg$^2$ in 56 pointings with filters (and $5\sigma$ limits) F062 (26.4), F087 (25.6), F106 (25.5), and F129 (25.4); while the narrow component covers an overlapping area of 5.6 deg$^2$ in 20 pointings with filters (and $5\sigma$ limits) F106 (26.7), F129 (26.6), F158 (26.5), and F184 (26.7)\footnote{For a definition of the \textit{Roman} filters see \url{https://roman.gsfc.nasa.gov/science/WFI_technical.html}}.

We use the TDE {\tt MOSFiT} models from \cite{Nicholl22} to generate a sample of 240,000 TDEs distributed uniformly across space (RA, DEC, and comoving distance), and time (throughout the 2 year duration of the HLTDS survey). Each TDE sample is generated from observations of a real TDE, which allows us to simply adopt the properties of existing TDEs without introducing additional parameter assumptions. {\tt MOSFiT} generates a blackbody SED at each epoch of a TDE's evolution, allowing us to convolve this SED with the \textit{Roman} passbands and measure their corresponding IR magnitudes. We then compare these magnitudes to the nominal $5\sigma$ limits of the HLTDS survey to quantify how many TDEs would be detected.

In Figure~\ref{fig:counts_Roman} we show the total number of TDEs per year expected to be found with at least 4 detections as part of the HLTDS. We adopt the volumetric TDE rate estimate from \cite{Velzen18}, which is defined as a function of TDE peak $g$-band luminosity as $\dot{N}_{0}  ~ (L / L_{0})^{a}$, where $\dot{N}_{0} = (1.9 \pm 0.7) \times 10^{-7}\, {\rm Mpc}^{-3}\,{\rm yr}^{-1}$ and $a=-1.6 \pm 0.2$ for $L_{0}=10^{43}\,{\rm erg}\,{\rm s}^{-1}$. Most TDEs in our sample have peak $g$-band luminosities of $\sim 10^{42}-10^{43}$ erg s$^{-1}$, which translates to a rate of $\sim 10^{-6} - 10^{-7}$ TDE Mpc$^{-3}$ yr$^{-1}$. Using these estimates we predict the \textit{Roman} HLTDS could find $\sim 300$ TDEs per year out to a redshift of $z \sim 7$ with at least 4 detections each, or $\sim 200$ ``well observed" TDEs with at least 20 detections spanning a minimum 70 day baseline between the first and last detection. It is also evident from Figure~\ref{fig:counts_Roman} that the distribution of TDEs found by the HLDTS peaks around a redshift of $z \sim 1 - 2$. 

We find that out of all TDEs injected into our simulation, $\sim 35$\% of the ones at $z < 0.2$ were detected with at least 4 data points in either the deep or wide components of the HLTDS. If we only consider the deep component of the HLTDS, we find that $\sim 81$\% of TDEs at $z < 0.2$ are detected with at least 4 data points. In the top panels of Figure~\ref{fig:lightcurves} we show some example light curves of TDEs detected by our simulation of the \textit{Roman} HLTDS. It is clear that while \textit{Roman} will find TDEs at high redshifts, the most distant ones at $z \sim 6$ will have poor coverage and are only detectable by the deep component of the HLTDS, which would make their characterization challenging. The wide component of the HLTDS can only detect TDEs out to a redshift of $z \sim 4$. Finding TDEs at redshifts this high will allow us to test different theories regarding the evolution of the TDE rate. For example, \cite{Kochanek16} predict that the rate of TDEs is expected to decrease at higher redshifts.

Given that \textit{Roman} does not have $g$- or $r$-band observations, we cannot directly run FLEET on the simulated \textit{Roman} light curves. Nevertheless, we can adopt the most frequent \textit{Roman} bands (F106 and F129) as proxies for $g$ and $r$, and run the late-time FLEET classifier on these. We show the fraction of TDEs recovered with \ptde$>0.5$ as a function of redshift in Figure~\ref{fig:fraction}. We determine that FLEET can recover with \ptde$>0.5$ about $10$\% of \textit{Roman} TDEs $z < 3.0$, or about $20$\% of TDEs at $z < 0.5$. This translates to $\sim 20$ TDEs that FLEET could uncover every year out to a redshift of $z = 0.5$, or $\sim 40$ TDEs out to a redshift of $z = 3.0$. We caution that these estimates assume FLEET would perform equally well on \textit{Roman} data as ZTF data. Eventually we will be able to re-train FLEET using \textit{Roman} data, including the host galaxy images from the Wide Area Survey, the light curves from the HLTDS, and the spectroscopic classifications for the $\sim 10$\% of detected transients that are expected to be produced by the \textit{Roman} slitless spectrograph \citep{Rose21}. But currently, our approximation depends on a number of factors. First, the difference in optical versus IR filters is likely to have a negative effect only for the closest TDEs at $z < 0.3$, since TDEs appear blue at early times they would be harder to distinguish from nuclear SNe in IR wavelengths. Nevertheless, this will likely not be a problem for the majority of TDEs discovered by \textit{Roman}, since the bulk of TDEs are expected to be detected at $z \sim 1 - 2$, which for the F129 and F158 filters correspond approximately to rest-frame $g$ and $r$ bands. The fact that we extrapolate a blackbody SED to calculate the magnitudes of the TDEs in the IR bands is possibly a conservative estimate given that \cite{Lu20} predict TDEs to be brighter in the IR than what a simple blackbody predicts. Additionally, \textit{Roman} will observe these transients with at least 4 filters, likely improving our selection criteria thanks to the doubling in bands available. Moreover, we do not account for the higher angular resolution of \textit{Roman}, which will likely increase our purity as it will allow us to better discern whether a transient is nuclear compared to what we can currently do with ZTF. Lastly, we do not account for intrinsic host-galaxy extinction of the transients from \textit{Roman}, which would only be different for the nearby TDEs observed in the rest-frame IR. In conclusion, our algorithms are likely to perform similarly well for high redshift TDEs, but it is uncertain how well they will perform for the closest TDEs with rest-frame IR observations. Obtaining a more realistic estimate of the number of TDEs that will be found by \textit{Roman} will require simulations of the \textit{Roman} survey, noise properties, images, and template subtraction capabilities and artifacts, and eventually real data from \textit{Roman} on which to train the FLEET algorithm; including the spectroscopic data for $\sim 10$\% of detected transients that are expected to be produced by the \textit{Roman} slitless spectrograph \citep{Rose21}.

Finally, we note that real-time follow-up of TDEs from \textit{Roman} will rely completely on the availability of real-time transient alerts. \textit{Therefore, we urge the implementation of a real-time transient alert system to be part of the \textit{Roman} survey.}

\subsection{\textit{Rubin}}\label{sec:lsst}

We estimate the number of TDEs that could be found by \textit{Rubin} \citep{LSST17} following the methods outlined in \cite{Villar18}. Here, we inject 30,000 TDE light curves into the 10 year \textit{Rubin} baseline simulation ({\tt baseline\_v2.2\_10yrs.db}), using the same methods for generating and injecting light curves outlined in \S\ref{sec:roman}. A smaller sample is used for \textit{Rubin} due to the computational limitations imposed by the more complex \textit{Rubin} survey simulation.

\begin{figure}
    \begin{center}
    \centering
    {\includegraphics[width=\columnwidth]{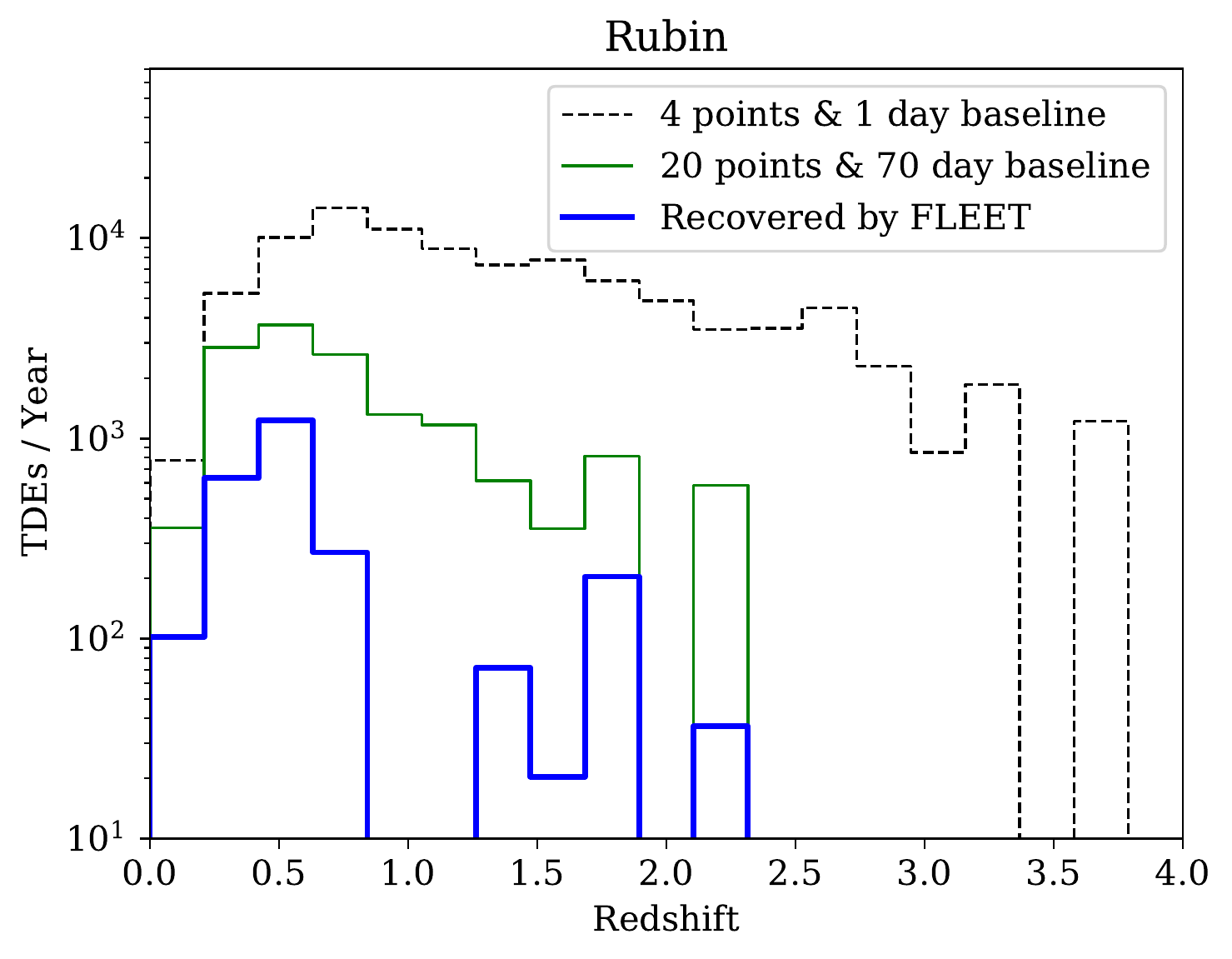}}
    \caption{Number of TDEs expected to be found by \textit{Rubin} per year, per redshift bin. The black dashed line shows the total number of TDEs with at least 4 detections spanning at least a 1 day baseline, while the green solid line shows only the ``well observed" TDEs with at least 20 detections over a 70 day baseline. The blue line shows the TDEs recovered by FLEET with \ptde$>0.5$. \label{fig:counts_LSST}}
    \end{center}
\end{figure}

In Figure~\ref{fig:counts_LSST} we show the number of TDEs expected to be found by \textit{Rubin} assuming the same volumetric rate from \cite{Velzen18} described in \S\ref{sec:roman}. We find that the number of TDEs detected by \textit{Rubin} does not extend past $z \sim 4$. We determine that $\sim 10^5$ TDEs could be detected by \textit{Rubin} every year with at least 4 data points each, or $\sim 10^4$ ``well observed" TDEs per year with at least 20 detections spanning a minimum 70 day baseline. This is comparable to the highest value of $\sim 8,000$ TDEs per year that \cite{Bricman} estimated would be discovered by \textit{Rubin}, which the authors calculate based on theoretical models of TDEs drawn from uniformly sampled impact factors and black hole masses, using a fixed rate of $10^{-5}$ TDEs per galaxy per year.

\begin{figure}
    \begin{center}
    \centering
    {\includegraphics[width=\columnwidth]{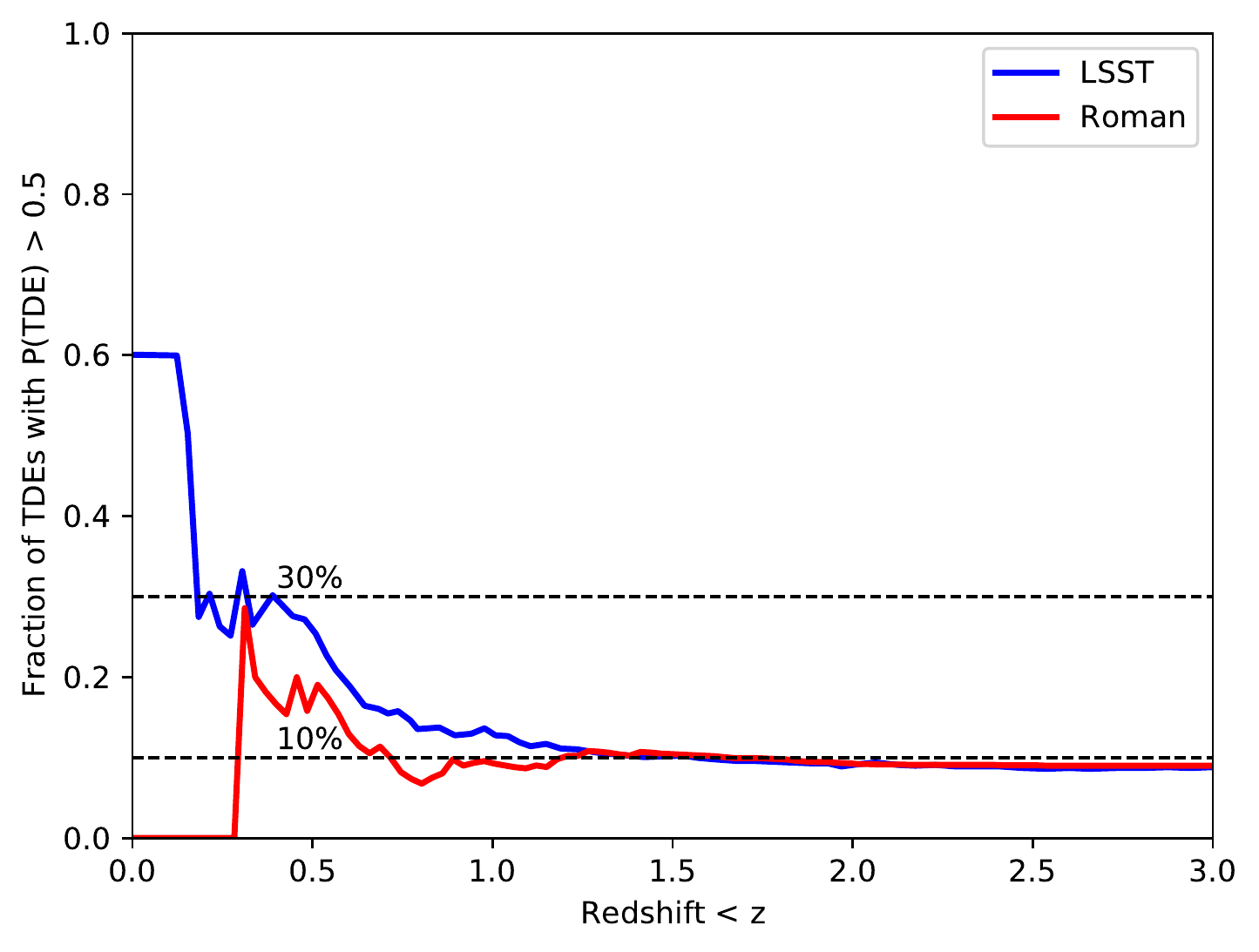}}
    \caption{Cumulative number of TDEs with \ptde$>0.5$ as a fraction of all TDEs lower than some redshift $z$ detected by \textit{Rubin} or \textit{Roman}. Out of all the injected TDEs detected in our \textit{Rubin} survey simulation, FLEET can recover $\sim 30$\% of the ones at a redshift $z < 0.5$, or $\sim 10$\% of the ones at $z < 1.5$. \label{fig:fraction}}
    \end{center}
\end{figure}

Since \textit{Rubin} includes $g$- and $r$-band observations, we can directly run FLEET on the light curves of the simulated TDEs that were detected by the survey. We run the late-time FLEET classifier on all the TDEs detected by the \textit{Rubin} simulation and show the fraction of TDEs with \ptde$>0.5$ as a function of redshift in Figure~\ref{fig:fraction}. We find that FLEET can recover with \ptde$>0.5$ about $10$\% of TDEs detected by \textit{Rubin} at a redshift $z < 1.5$, or about $30$\% of TDEs at a redshift $z < 0.5$. This translates to $\sim 2,000$ TDEs that FLEET could uncover every year out to a redshift of $z = 0.5$, or $\sim 3,000$ TDEs out to a redshift of $z = 1.5$. We show sample light curves of well observed TDEs from the \textit{Rubin} simulation located at redshifts of $z = 0.1$, 0.5, and 1.5 in the bottom row of Figure~\ref{fig:lightcurves}. We note that these estimates come from running the current version of FLEET, trained on ZTF data, on \textit{Rubin} data. Optimizing the algorithm with \textit{Rubin} data is only expected to improve these estimates. Training the algorithm on $u$-band will be particularly useful, since TDEs are known to separate well from other transients in this band \citep{Velzen20_TDEs}.

Despite the fact that \textit{Rubin} is expected to find a larger total number of TDEs than \textit{Roman}, the fraction of TDEs recovered by \textit{Rubin} with 20 data points is an order of magnitude lower than the fraction of TDEs recovered with 4 data points. Meaning \textit{Rubin} is likely to find many TDEs that will go unclassified due to their poor light curve coverage. This is not the case for the \textit{Roman} HLTDS survey, for which the fraction of TDEs recovered with 4 data points is almost the same as the fraction of TDEs recovered with 20 data points, as a consequence of the small footprint, cadence, and depth of the \textit{Roman} HLTDS. 

\section{Conclusions}
\label{sec:conclusions}

We have presented a new version of FLEET, a machine learning classifier designed specifically to rapidly identify TDEs with a high purity, and without the need for redshift information. We trained this classifier on a sample of 4,779 spectroscopically classified transients, including 45 TDEs. We provide two classifiers, a rapid one trained on 20 days of photometry meant to be used for real-time classification, and a late-time classifier trained on 40 days of data meant to be used for more robust estimates even if a transient has begun to fade. Both classifiers use light curve and contextual host galaxy information to calculate the probability of a transient to be a TDE, \ptde. Our key findings are:

\begin{itemize}

\item The most important features for distinguishing TDEs from other transients are the normalized host separation $R_n$ and the light curve color during peak $(g-r)_P$.

\item We can recover TDEs with a purity of $\approx 30$\% using the rapid classifier for events with \ptde$>0.5$. This is a factor of $\sim 60$ improvement compared to random selection. The corresponding completeness for this threshold is $\approx 40$\%, or about 5 times better than using simple selection cuts.

\item We find a peak purity of $\approx 50\%$ can be achieved with the rapid classifier for transients with \ptde$>0.8$, corresponding to a completeness of $\approx 20$\%.

\item The late-time classifier trained on 40 days of data performs similarly with a completeness of $\approx 30\%$ for transients with \ptde$>0.8$, but with a higher peak purity of $\approx 90$\%.

\end{itemize}

Additionally, we explore the application of FLEET to current and future time-domain surveys. We present a list of 39 TDE candidates with \ptde$> 0.5$ that were found by ZTF but remain currently unclassified. We inject TDE light curves into simulations of the \textit{Roman} and \textit{Rubin} time-domain surveys to estimate how many TDEs can be recovered given their current survey designs. We find that using FLEET we could recover $\sim 30$ TDEs per year with the \textit{Roman} HLTDS out to a redshift of $z \sim 7$, and $\sim 3,000$ TDEs per year with \textit{Rubin} out to a redshift of $z \sim 3$. Showing that tools like FLEET will be not only useful, but necessary, in the era of big data and large time-domain surveys. Given the successful performance of FLEET up to date, the predictions for detecting TDEs with \textit{Rubin} and the lack of strong biases against physical parameters, we are confident we can begin running FLEET on \textit{Rubin} data and discovering interesting transients as soon as the survey begins.

\acknowledgements

S.G. is supported by an STScI Postdoctoral Fellowship. The Berger Time Domain group at Harvard is supported in part by NSF and NASA grants, including support by the NSF under grant AST-2108531, as well as by the NSF under Cooperative Agreement PHY-2019786 (The NSF AI Institute for Artificial Intelligence and Fundamental Interactions \url{http://iafi.org/}). M.N. is supported by the European Research Council (ERC) under the European Union’s Horizon 2020 research and innovation programme (grant agreement No.~948381) and by a Fellowship from the Alan Turing Institute. Operation of the Pan-STARRS1 telescope is supported by the National Aeronautics and Space Administration under grant No. NNX12AR65G and grant No. NNX14AM74G issued through the NEO Observation Program. This work has made use of data from the European Space Agency (ESA) mission {\it Gaia} (\url{https://www.cosmos.esa.int/gaia}), processed by the {\it Gaia} Data Processing and Analysis Consortium (DPAC, \url{https://www.cosmos.esa.int/web/gaia/dpac/consortium}). Funding for the DPAC has been provided by national institutions, in particular the institutions participating in the {\it Gaia} Multilateral Agreement. This research has made use of NASA’s Astrophysics Data System. This research has made use of the SIMBAD database, operated at CDS, Strasbourg, France. Based on observations obtained with MegaPrime/MegaCam, a joint project of CFHT and CEA/IRFU, at the Canada-France-Hawaii Telescope (CFHT) which is operated by the National Research Council (NRC) of Canada, the Institut National des Science de l'Univers of the Centre National de la Recherche Scientifique (CNRS) of France, and the University of Hawaii. This work is based in part on data products produced at Terapix available at the Canadian Astronomy Data Centre as part of the Canada-France-Hawaii Telescope Legacy Survey, a collaborative project of NRC and CNRS. This research has made use of the NASA/IPAC Extragalactic Database, which is funded by the National Aeronautics and Space Administration and operated by the California Institute of Technology.

\facilities{ADS, TNS}
\software{Astropy \citep{astropy}, MOSFiT \citep{mosfit}, extinction (\citep{Barbary16}), Matplotlib \citep{matplotlib}, emcee\citep{foreman13}, NumPy \citep{numpy}, scikit-learn \citep{Pedregosa12}, SMOTE \citealt{Chawla02}, FLEET \citep{Gomez20_FLEET}}

\bibliography{references}

\appendix
\setcounter{figure}{0}
\setcounter{table}{0}

In Table~A.\ref{tab:TDEs} we include the list of all TDEs used to train the FLEET classifier. Only TDEs with at least 2 $g$-band and 2 $r$-band points are included. We exclude AT\,2019eve from our classifier because even though it was originally classified as a TDE \citep{Velzen20}, the classification has since been retracted due to the uncertainty in its spectral features \citep{Hammerstein22}. Similarly, we exclude AT\,2018dyk since it was reclassified from a TDE to a changing look AGN by \cite{Frederick19}.

\startlongtable
\begin{deluxetable*}{ccc|ccc|ccc}
    \tablecaption{TDEs Used for Training FLEET \label{tab:TDEs}}
    \tablehead{\colhead{Name} & \colhead{Redshift}  & \colhead{Reference} & \colhead{Name} & \colhead{Redshift}  & \colhead{Reference} & \colhead{Name} & \colhead{Redshift}  & \colhead{Reference}}
    \startdata
    AT\,2018bsi  & 0.0510 & [1,2]       & AT\,2019meg  & 0.1520 & [1,3]       & AT\,2020vwl  & 0.0350 & [8]    \\
    AT\,2018hco  & 0.0900 & [3,1]       & AT\,2019mha  & 0.1480 & [1,3]       & AT\,2020wey  & 0.0273 & [1]    \\
    AT\,2018hyz  & 0.0458 & [1,3,21,22] & AT\,2019qiz  & 0.0151 & [1,3,25,27] & AT\,2020ysg  & 0.2770 & [1]    \\
    AT\,2018iih  & 0.2120 & [1,3]       & AT\,2019teq  & 0.0878 & [1]         & AT\,2020zso  & 0.0610 & [1,26] \\
    AT\,2018jbv  & 0.3400 & [1]         & AT\,2019vcb  & 0.0890 & [1]         & AT\,2021ack  & 0.1330 & [9]    \\
    AT\,2018lna  & 0.0910 & [1,3]       & AT\,2020ddv  & 0.1600 & [1]         & AT\,2021axu  & 0.1900 & [10]   \\
    AT\,2018lni  & 0.1380 & [1]         & AT\,2020mbq  & 0.0930 & [1]         & AT\,2021ehb  & 0.0170 & [11]   \\
    AT\,2018zr   & 0.0710 & [1,3]       & AT\,2020mot  & 0.0700 & [1]         & AT\,2021gje  & 0.3580 & [12]   \\
    AT\,2019azh  & 0.0223 & [1,3,23]    & AT\,2020neh  & 0.0620 & [6]         & AT\,2021jjm  & 0.1530 & [13]   \\
    AT\,2019bhf  & 0.1206 & [1,3]       & AT\,2020nov  & 0.0840 & [7]         & AT\,2021jsg  & 0.1260 & [14]   \\
    AT\,2019cho  & 0.1930 & [1,3]       & AT\,2020ocn  & 0.0700 & [1]         & AT\,2021mhg  & 0.0730 & [15]   \\
    AT\,2019dsg  & 0.0512 & [1,3,24]    & AT\,2020opy  & 0.1590 & [1]         & AT\,2021nwa  & 0.0470 & [16]   \\
    AT\,2019ehz  & 0.0740 & [1,3]       & AT\,2020pj   & 0.0680 & [1]         & AT\,2021sdu  & 0.0590 & [18]   \\
    AT\,2019gte  & 0.0860 & [4]         & AT\,2020qhs  & 0.3450 & [1]         & AT\,2021uqv  & 0.1060 & [19]   \\
    AT\,2019lwu  & 0.1170 & [1,3]       & AT\,2020riz  & 0.4350 & [1]         & AT\,2021yte  & 0.0530 & [20]   \\
    \enddata
    \tablecomments{All TDEs used to train our classifiers, listed alphabetically. 1: \cite{Hammerstein22}; 2: \cite{Velzen20}; 3: \cite{Nicholl22}; 4: \cite{Swann_2019gte}; 6: \cite{Dahiwale_2020neh}; 7: \cite{Dahiwale_2020nov}; 8: \cite{Hammerstein_2020vwl}; 9: \cite{Hammerstein_2021ack}; 10: \cite{Hammerstein_2021axu}; 11: \cite{Yao_2021ehb}; 12: \cite{Hammerstein_2021gje}; 13: \cite{Yao_2021jjm}; 14: \cite{Yao_2021jsg}; 15: \cite{Chu_2021mhg}; 16: \cite{Yao_2021nwa}; 18: \cite{Chu_2021sdu}; 19: \cite{Yao_2021uqv}; 20: \cite{Yao_2021yte}; 21: \cite{Short20}; 22: \cite{Gomez20_hyz}; 23: \cite{Liu22}; 24: \cite{Cannizzaro21}; 25: \cite{Hung21}; 26: \cite{Wevers22}; 27: \cite{Nicholl20}}
\end{deluxetable*}

\end{document}

%% file: affil.tex
\newcommand{\STScI}{\affiliation{Space Telescope Science Institute, 3700 San Martin Dr, Baltimore, MD 21218, USA}}

\newcommand{\Einstein}{\altaffiliation{Einstein Fellow}}

\newcommand{\CfA}{\affiliation{Center for Astrophysics \textbar{} Harvard \& Smithsonian, 60 Garden Street, Cambridge, MA 02138-1516, USA}}

\newcommand{\Birmingham}{\affiliation{Birmingham Institute for Gravitational Wave Astronomy and School of Physics and Astronomy, University of Birmingham, Birmingham B15 2TT, UK}}
\newcommand{\CIERA}{\affiliation{Center for Interdisciplinary Exploration and Research in Astrophysics and Department of Physics and Astronomy, \\Northwestern University, 1800 Sherman Ave., 8th Floor, Evanston, IL 60201, USA}}

\newcommand{\Leiden}{\affiliation{Leiden Observatory, Leiden University, PO Box 9513, 2300 RA Leiden, The Netherlands}}
\newcommand{\PSUa}{\affiliation{Department of Astronomy \& Astrophysics, The Pennsylvania State University, University Park, PA 16802, USA}}
\newcommand{\PSUb}{\affiliation{Institute for Computational \& Data Sciences, The Pennsylvania State University, University Park, PA 16802, USA}}
\newcommand{\PSUc}{\affiliation{Institute for Gravitation and the Cosmos, The Pennsylvania State University, University Park, PA 16802, USA}}
\newcommand{\IAIFI}{\affiliation{The NSF AI Institute for Artificial Intelligence and Fundamental Interactions}}